\documentclass[review]{elsarticle}

\usepackage{lineno,hyperref}

\usepackage{latexsym,amsthm,amsmath,amssymb,mathrsfs,dsfont,upgreek,stackengine,graphicx,caption, float,enumerate}
\usepackage{pdfpages,courier,subcaption,inputenc,tikz,verbatim,blkarray, listings,latexsym,pythonhighlight,xcolor,geometry,xparse}
\usepackage{algorithmic}
\usepackage{multirow, array, textcomp, subcaption, stmaryrd}

\newcolumntype{P}[1]{>{\centering\arraybackslash}p{#1}} % To make the data in the table be centering
\newcolumntype{M}[1]{>{\centering\arraybackslash}m{#1}}

\usepackage[ruled,vlined]{algorithm2e}

\newtheorem{remark}{Remark}

\modulolinenumbers[5]

\journal{Journal of Computational Physics}

\begin{document}

\begin{frontmatter}

\title{Numerical aspects of Casimir energy computation in acoustic scattering}
% \tnotetext[tnote1]{This is an example for title footnote coding.}
\tnotetext[mytitlenote]{Supported by Leverhulme grant RPG-2017-329.}

%\tnotetext[mytitlenote]{Fully 	documented templates are available in the elsarticle package on \href{http://www.ctan.org/tex-archive/macros/latex/contrib/elsarticle}{CTAN}.}

%% Group authors per affiliation:
\author[mymainaddress]{Xiaoshu Sun\corref{mycorrespondingauthor}}
\cortext[mycorrespondingauthor]{Corresponding author}
\ead{xiaoshu.sun.18@ucl.ac.uk}

\author[mymainaddress]{Timo Betcke}
\ead{t.betcke@ucl.ac.uk}
\address[mymainaddress]{Department of Mathematics, University College London, London, WC1E 6BT, UK}

\author[mysecondaryaddress]{Alexander Strohmaier}
\ead{a.strohmaier@leeds.ac.uk}
\address[mysecondaryaddress]{School of Mathematics, University of Leeds, Leeds, LS2 9JT, UK}

%% or include affiliations in footnotes:

\begin{abstract}
    Computing the Casimir force and energy between objects is a classical problem of quantum theory going back to the 1940s. 
    Several different approaches have been developed in the literature often based on different physical principles. Most notably a representation 
    of the Casimir energy in terms of determinants of boundary layer operators makes it accessible to a numerical approach. In this paper, we first give 
    an overview of the various methods and discuss the connection to the Krein-spectral shift function and computational aspects.
    We propose variants of Krylov subspace methods for the computation of the Casimir energy for large-scale problems and demonstrate Casimir computations 
    for several complex configurations. This allows for Casimir energy calculation for large-scale practical problems and significantly speeds up the computations in that 
    case.
   \end{abstract}

\begin{keyword}
Krein spectral shift function \sep Casimir energy \sep Krylov subspace \sep inverse-free generalized eigenvalue problem \sep Bempp-cl
\end{keyword}

\end{frontmatter}

\section{Introduction}\label{Introduction}
Casimir interactions are forces between objects such as perfect conductors. Hendrik Casimir predicted and computed this effect in the special case of two planar 
conductors in 1948 using a divergent formula for the zero point energy and applying regularisation to it \cite{casimir1948attraction}. This resulted in the 
famous formula for the attractive Casimir force per unit area 
\begin{align*}
    F(a) = -\frac{1}{A}\frac{\partial \mathcal{E}}{\partial a} = -\frac{\hbar c\pi^{2}}{240a^{4}},
\end{align*}
between two perfectly conducting plates, where $A$ is the cross-sectional area of the boundary plates and $\mathcal{E}$ is the Casimir energy as 
computed from a zeta regularised mode sum. The result here is for the electromagnetic field which differs by a factor of two from the force resulting from a 
massless scalar field.
This force was measured experimentally by Sparnaay 
about 10 years later \cite{sparnaay1958measurements} and the Casimir effect has since become famous for its intriguing derivation and its counterintuitive nature.
In 1996, precision measurements of the Casimir force between 
 extended bodies were conducted by S.K. Lamoreaux \cite{lamoreaux1997demonstration}  confirming the theoretical predictions including corrections for 
realistic materials. From 2000 to 2008, the Casimir force has been measured in various 
experimental configurations, such as cylinder-cylinder \cite{ederth2000template}, plate-plate \cite{bressi2002measurement}, 
sphere-plate \cite{krause2007experimental} and sphere-comb \cite{chan2008measurement}. 
The presence of the Casimir force has also been quoted as evidence for the zero point energy of the vacuum having direct physical significance.
The classical way to compute Casimir forces mimicks Casimir's original computation and is based on zeta function regularisation of the vacuum energy. 
This has been carried out for a number of particular geometric situations (see \cite{bordag2001new, bordag2009advances, elizalde1989expressions, elizalde1990heat, kirsten2001spectral} and references therein). 
The derivations are usually based on special functions and their properties and require explicit knowledge of the spectrum of the Laplace operator.

In 1960s, Lifshitz and collaborators extended and modified this theory to the case of dielectric media \cite{dzyaloshinskii1961general} 
gave derivations based on the stress energy tensor. It has also been realised by quantum field theorists  
(see e.g. \cite{brown1969vacuum, dzyaloshinskii1961general, deutsch1979boundary, kay1979casimir, scharf1992casimir}) with various degrees of mathematical rigour that 
the stress energy approach yields Casimir's formula directly without the need for renormalisation or artificial regularisation.
This tensor is defined by comparing the induced vacuum states of the quantum field with boundary conditions and the free theory. 
Once the renormalised stress energy tensor is mathematically defined, the computation of the Casimir energy density becomes a problem of spectral 
geometry (see e.g. \cite{fulling2007vacuum}). The renormalised stress energy tensor and its relation to the Casimir effect can be understood at the 
level of rigour of axiomatic algebraic quantum field theory. We note however that the computation of the local energy density is non-local and requires 
some knowledge of the spectral resolution of the Laplace operator, the corresponding problem of numerical analysis is therefore extremely hard.

Lifshitz and collaborators also offered an alternative description based on the van der Waals forces between molecules.
The plates consist of a collection of atomic-scale electric dipoles randomly oriented in the absence of the external forcing field. Quantum and thermal 
fluctuations may make the dipoles align spontaneously, resulting in a net electric dipole moment. The dipoles in the opposite plate feel this field
across the gap and align as well. The two net electric dipole moments make the two plates attract each other. This approach emphasizes the 
influence from the materials more than the fluctuations in the empty space between the plates.

Somewhat independently from the spectral approach determinant formulae based on the van der Waal's mechanism were derived by various authors. 
We note here Renne \cite{renne1971microscopic} who gives a determinant formula for the van der Waals force based on microscopic considerations. Various authors 
found path-integral derivations of Casimir forces based on considerations of surface current fluctuations 
\cite{bimonte2017nonequilibrium, emig2007casimir, emig2006casimir, EGJK2008, emig2008casimir, kenneth2006opposites, kenneth2008casimir, milton2008multiple, rahi2009scattering}. The final formulae proved suitable for numerical schemes 
and were also very useful to obtain asymptotic formulae for Casimir forces for large and small separations. The mathematical relation between the various 
approaches remained unclear, with proofs of equality only available in special cases.
A full mathematical justification of the determinant formulae as the trace of an operator describing the Casimir energy was only recently achieved in 
\cite{MR4484208} for the scalar field and \cite{strohmaier2021classical} for the electromagnetic field. It was also proved recently in \cite{fang2021mathematical} that the 
formulae arising from the stress energy tensor and from the determinant formulae give the same Casimir forces.

The precise mathematical formulation for the Casimir energy computations in this framework is closely related to the method of boundary layer operators, a well established theory to deal with computational acoustic and electrodynamic wave dynamic.
It is particularly important for numerical investigations that the involved quantities are well defined on the correct function spaces, and that an approximation theory exists for them.
We will therefore describe in some detail the precise mathematical setting and review the theory from this point of view.

Let $\Omega \subset \mathbb{R}^{d}$ be a non-empty bounded open subset with Lipschitz boundary $\partial \Omega$, which is the union of connected open 
sets $\Omega_{j}$, for $j = 1, \dots, N$. We assume that the complement $\mathbb{R}^{d} \backslash \Omega$ of $\Omega$ is connected and the closures of $\Omega_{j}$ 
are pairwise non-intersecting.
We denote the  $N$ connected components of the boundary $\partial\Omega$ by $\partial\Omega_{j}$. 
We will think of the open set $\Omega$ as a collection of objects $\Omega_{j}$ placed in $\mathbb{R}^{d}$ and will refer to them as {\sl obstacles}.

Then, several unbounded self-adjoint operators densely defined in $L^{2}(\mathbb{R}^{d})$
can be defined.
\begin{itemize}
    \item The operator $\Delta$ is the Laplace operator with Dirichlet boundary conditions on $\partial\Omega$.
    \item For $j = 1, \dots, N$, the operator $\Delta_{j}$ is the Laplace operator with Dirichlet boundary conditions on $\partial\Omega_{j}$.
    \item The operator $\Delta_{0}$ is the ``free'' Laplace operator on $\mathbb{R}^{d}$ with domain $H^{2}(\mathbb{R}^{d})$.
\end{itemize}

These operators contain the dense set $C^\infty_0(\mathbb{R}^d \setminus \partial \Omega)$ in their domains.
If $f: \mathbb{R} \to \mathbb{R}$ is a polynomially bounded function this set is also contained in the domain of the operators
$f(\Delta^{\frac{1}{2}}), f(\Delta_{j}^{\frac{1}{2}})$, and $f(\Delta_{0}^{\frac{1}{2}})$, in particular the operator
$$
 D_{f} = f(\Delta^{\frac{1}{2}}) - f(\Delta_{0}^{\frac{1}{2}}) - \left(\sum_{j = 1}^{N}[f(\Delta_{j}^{\frac{1}{2}}) - f(\Delta_{0}^{\frac{1}{2}})]\right)
$$
is densely defined. It was shown in \cite{MR4484208} that under additional analyticity assumptions on $f$ the operator
$D_{f}$ is bounded and extends by continuity to a trace-class operator on $L^2(\mathbb{R}^{d})$. 
These analyticity assumptions are in particular satisfied by $f(k) = (k^{2}+ m^{2})^{\frac{s}{2}}$ for any $s > 0, m \geq 0$ and one has
\begin{align}\label{trace formula in terms of the boundary op}
    \text{Tr}\left( D_{f} \right)  = \frac{s}{\pi} \sin\left(\frac{\pi}{2} s\right) \int_{m}^{\infty} k (k^{2} + m^{2})^{\frac{s}{2}-1}\Xi(\mathrm{i} k) dk,
\end{align}
where the function $\Xi$ is given by
$$
 \Xi(k) = \log \det V_{k} \tilde V_{k}^{-1}
$$
and the operators $V_{k}$ and $\tilde V_{k}$ are certain single layer operators that will be defined later. 
It was proved in  \cite{MR4484208} that the above determinant is well-defined in the sense of Fredholm as the operator $V_{k} \tilde V_{k}^{-1}$ near the positive imaginary axis differs
from the identity operator by a trace-class operator on the Sobolev space $H^\frac{1}{2}(\partial \Omega)$.
We remark here that the paper \cite{MR4484208} assumed the boundary to be smooth and the operators $V_k  \tilde V_k^{-1}$ was considered as a 
map on $L^2(\partial \Omega)$. The main result of the paper also holds for Lipschitz boundaries if $L^2(\partial \Omega)$ is replaced by $H^\frac{1}{2}(\partial \Omega)$. This requires minor modifications of the proof but we will not discuss this further here, as we are now focusing on computational aspects. The Casimir energy for the scalar field of mass $m\geq0$ is then given by $\frac{\hbar c}{2}\text{Tr}\left( D_{f} \right)$. In particular for the massless scalar field one thus obtains
$$
\zeta = \frac{\hbar c}{2 \pi} \int _{0}^{\infty} \Xi(\mathrm{i}k) dk.
$$
This formula is equivalent to expressions that have appeared in the physics literature and has since emerged as an efficient tool to compute Casimir forces between compact objects.

We also recall that by the Birman-Krein formula we have for any even function $h \in \mathcal{S}(\mathbb{R})$ the equality
\begin{align}\label{B-K formula}
    \text{Tr}\left(h(\Delta^{\frac{1}{2}}) - h(\Delta_{0}^{\frac{1}{2}}) - \left(\sum_{j = 1}^{N}[h(\Delta_{j}^{\frac{1}{2}}) - h(\Delta_{0}^{\frac{1}{2}})]\right)\right)  = \int_{0}^{\infty}h'(k)\xi(k)dk,
\end{align}
where 
\begin{align*}
    \xi(k) = \frac{1}{2\pi \mathrm{i}}\log\left(\frac{\det(S(k))}{\det(S_{1,k})\cdots\det(S_{N,k}(k))}\right)
\end{align*}
will be called the relative Krein spectral shift function. Here, $S_{j,k}$ are the scattering matrices of $\Delta_{j}$ associated to the objects $\Omega_{j}$. Note here that the class of functions for which this is true can be relaxed to a certain extent, but even the most general version does not allow unbounded functions such as $f(k)$ with $s>0, m\geq 0$.
The function $\Xi(k)$ can however be related via a Laplace transform to the Fourier transform of the relative spectral shift function (see \cite{MR4396069}). Under mild convexity assumptions this can be connected to the Duistermaat-Guillemin trace formula in obstacle scattering theory to give an asymptotic expansion of  $\Xi(k)$ 
in terms of the minimal distance $\delta>0$ between the obstacles and the linearised Poincar\'e map of the bouncing ball orbits between the obstacles of that length. One has
$$
 \Xi(k) =- \sum_{j} \frac{1}{|\det(I - P_{\gamma_j})|^{\frac{1}{2}}} e^{2 i \delta k} + o(e^{- 2 \delta \text{Im}{k}}),
$$
where the sum is over  bouncing ball modes of length $2 \delta$ and $P_{\gamma_j}$ is the associated Poincar\'e map, where $\gamma_{j}$ is the shortest bouncing ball orbits. We note here that the convexity assumption is needed here only for the precise asymptotic, but 
exponential decay at rate $e^{2 i \delta k}$ holds independent of that (see \cite{MR4396069}).
%The Casimir energy of the configuration $\Omega$ for a massless scalar field would then be given by $D_f$ in the case when $f(k)=k$ and is therefore equal to
%$$
%\zeta = \frac{\hbar c}{2 \pi} \int _{0}^{\infty} \Xi(\mathrm{i}k) dk.
%$$

The approach via determinants of boundary layer operators gives a numerical framework for computing the Casimir energy for the massless scalar field \footnote{The mathematical theories and numerical experiments in the Maxwell case have been done as well and they will be 
reported in another paper.} which we describe again in somewhat more detail in Section \ref{Numerical methods for computing the Casimir energy}. Using the determinant formula as a starting point 
two efficient methods for computing the integrand of the Casimir energy will be illustrated in Section \ref{Krylov subspace for generalized eigenvalue problem}
which allows us compute large-scale problems. In Section \ref{Numerical experiments}, several examples on computing the Casimir energy between 
 compact objects will be shown and we will also compare our results with others computed in other methods. Note that all the tests and examples in this paper were computed 
with version 0.2.4 of the Bempp-cl library \cite{scroggs2017software}. Finally, Section \ref{Conclusion} will conclude 
our paper and discuss the future plan as well.

% and discuss the spectral properties 
% of the block matrices constructed from the integral operators in Section \ref{Spectral property of the integral operators}.

\section{Numerical methods for computing the Casimir energy in acoustic scattering}\label{Numerical methods for computing the Casimir energy}
% !TEX root =  main.tex

In this section, we give details of computing the Casimir energy via boundary integral operator discretisations. 
Assume 
$\Omega^{-}\subset \mathbb{R}^{d}$, for $d \geq 2$ is the interior open bounded domain that the scatterer occupies with piecewise smooth Lipschitz boundary $\Gamma$. The exterior domain is denoted as 
$\Omega^{+} = \mathbb{R}^{d}\backslash\overline{\Omega^{-}}$. $\boldsymbol{n}$ is the almost everywhere defined exterior unit normal to the surface $\Gamma$ pointing outwards from $\Omega^{-}$ and 
$\boldsymbol{n}_{\boldsymbol{x}}$ is normal to $\Gamma$ at the point $\boldsymbol{x}\in\Gamma$.

In the scalar case, the Casimir energy can be expressed in terms of certain single-layer boundary operator, which we will define below. We then present its relationship with the Krein-Spectral shift function and demonstrate how it can practically be computed.

\subsection{The single-layer boundary operator}
For the bounded interior domain $\Omega^{-}$ or the unbounded exterior domain $\Omega^{+}$, the space of the (locally) square integrable functions is 
\begin{align*}
    L^{2}(\Omega^{-}) &:= \left\{f:\Omega^{-}\rightarrow\mathbb{C}, f \text{ is Lebesgue measurable and} \int_{\Omega^{-}}|f|^{2} < \infty \right\},\\
    L_{\text{loc}}^{2}(\Omega^{+}) &:= \left\{f:\Omega^{+}\rightarrow\mathbb{C},\ f \text{ is Lebesgue measurable and} \int_{K}|f|^{2} < \infty, \ \text{for all compact}\ K \subset \overline{\Omega^{+}} \right\}
\end{align*}
and note that the subscript ``loc'' can be removed if the domain is bounded (i.e. $L_{\text{loc}}^{2}(\Omega^{-}) = L^{2}(\Omega^{-})$).
We denote by $H_{\text{loc}}^{s}(\Omega^{\pm})$ the standard Sobolev spaces associated with the Lipschitz domains. In particular, for integers $s\geq 0$, we have 
\begin{align*}
    H_{\text{loc}}^{s}(\Omega^{\pm}):=\left\{f\in L_{\text{loc}}^{2}(\Omega^{\pm}), \forall\alpha \text{ s.t.} |\alpha|\leq s, D^{\alpha}f\in L_{\text{loc}}^{2}(\Omega^{\pm})\right\},
\end{align*}
where $\alpha = (\alpha_{1}, \alpha_{2}, \dots, \alpha_{d})$ is a multi-index and $|\alpha| = \alpha_{1} + \alpha_{2} + \dots + \alpha_{d}$, and 
the derivative is defined in the weak sense.
One also has the Sobolev spaces on the boundary $H^{s}(\Gamma)$ for any $-\frac{1}{2} \leq s \leq \frac{1}{2}$.
For a function $p$ on $\Omega$ that is continuous on the boundary we have the trace map $\gamma_{D}^{\pm}$ defined by
\begin{align*}
    \gamma_{\text{D}}^{\pm}p(\boldsymbol{x}):=\lim_{\Omega^{\pm}\ni\boldsymbol{x'}\rightarrow\boldsymbol{x}\in\Gamma}p(\boldsymbol{x'})
\end{align*}
that maps the function to its boundary value. This trace map is well-known to extend continuously to a map
$\gamma_{D}^{\pm}:  H_{\text{loc}}^1(\Omega) \to H^{1/2}(\Gamma) $. For the purposes of this paper it is sufficient to understand $H^{1/2}(\Gamma)$ as range space of the trace operator on $H_{\text{loc}}^1(\Omega)$ . We also need the space $H^{-1/2}(\Gamma)$, which is the dual space of $H^{1/2}(\Gamma)$ with $L^2(\Gamma)$ as pivot space.

We can now define the single-layer boundary $V_{k}:H^{-1/2}(\Gamma)\rightarrow H^{1/2}(\Gamma)$ as the continuous extension of the map defined in terms of an integral kernel as follows

\begin{align*}
    (V_{k}\mu)(\boldsymbol{x}) := \int_{\Gamma}g_{k}(\boldsymbol{x},\boldsymbol{y})\psi(\boldsymbol{y})dS_{\boldsymbol{y}}, \ \ \ \ \ 
    \text{for}\ \mu\in H^{-\frac{1}{2}}(\Gamma) \  \text{and} \ \boldsymbol{x}\in\Gamma.
\end{align*}
Here, 
\begin{align}\label{Green's function}
    g_{k}(\boldsymbol{x},\boldsymbol{y}) = \begin{cases}
          \frac{\mathrm{i}}{4}H_{0}^{(1)}(k|\boldsymbol{x}-\boldsymbol{y}|), \ \ \ \ &\text{for} \ d = 2\\
          \frac{e^{ik|\boldsymbol{x}-\boldsymbol{y}|}}{4\pi|\boldsymbol{x} - \boldsymbol{y}|}, \ \ \ \ &\text{for} \ d = 3,
        \end{cases}
\end{align}
with $H_{0}^{(1)}$  a Hankel function of the first kind.

\subsection{The formula of the Casimir energy}
% By \cite{MR4484208}, the Krein spectral shift function is defined as 
% \begin{align*}
%     \xi(k) = \frac{1}{2\pi \mathrm{i}}\log\left(\frac{\det(S(k))}{\det(S_{1,k})\cdots\det(S_{N,k})}\right),
% \end{align*}
% where $S_{i,n}$ is the scattering matrix associated with the $n$th scatterer. These scattering matrices can be constructed  $S_{i,n} = I + 2T_{i,n}$, where 
% $I$ is the identity matrix and $T_{i,n}$ is the $T$-matrix. The method of computing the $T$-matrix is fully discussed in \cite{waterman1969new} and 
% \cite{ganesh2008far}.

    Consider $\Omega$ as a domain assembled from individual objects $\Omega_{i}$ as described before. Let $V_{k}$ be the single-layer boundary operator defined on the boundary 
    $\partial\Omega = \bigcup_{i = 1}^{N}\partial\Omega_{i}$, and $\tilde{V}_{k}$ is the ``diagonal part'' of $V_{k}$ by restricting the integral 
    kernel to the subset $\bigcup_{i = 1}^{N}\partial\Omega_{i}\times\partial\Omega_{i}\subset\partial\Omega\times\partial\Omega$. Then the operator 
    $V_{k}\tilde{V}_{k}^{-1} - I$ with $I$ the identity operator is trace-class (\cite{MR4484208}) and one can therefore define the function
    \begin{align*}
        \Xi(k) = \log\det\left(V_{k}\tilde{V}_{k}^{-1}\right),
    \end{align*}
    using the Fredholm determinant $\det(V_{k}\tilde{V}_{k}^{-1})$
    \footnote{The Fredholm determinant is a generalization of a determinant of finite dimensional matrix to finite dimensional linear operator 
    which differ from the identity operator by a trace class operator \cite[Section 6.5.2]{MR2300779}. Since the operator $V_{k}\tilde{V}_{k}^{-1} - I$ 
    with $I$ the identity operator is trace-class in the close upper half space \cite[Theorem 1.7]{MR4484208}, the determinant $\det(V_{k}\tilde{V}_{k}^{-1})$ is well-defined.}.
  One has (taking $m = 0$ and $s = 1$ in \eqref{trace formula in terms of the boundary op})
    \begin{align}\label{slp and matrix}
        \emph{\text{Tr}}\left(\Delta^{\frac{1}{2}} + (N - 1)\Delta_{0}^{\frac{1}{2}} - \sum_{i = 1}^{N}\Delta_{j}^{\frac{1}{2}}\right)  =  \frac{1}{\pi}\int_{0}^{\infty}\Xi(\mathrm{i}k)dk.
    \end{align}

As explained in the introduction \eqref{slp and matrix} is used to compute the Casimir energy between the objects and the formula can be written as
\begin{align}\label{KSSF and CasE}
    \zeta = \frac{\hbar c}{2\pi}\int_{0}^{\infty}\Xi(\mathrm{i}k)dk.
\end{align}

\begin{remark}
    There is a relation between the relative Krein spectral shift function and the single-layer boundary integral operator. That is,
    for $k > 0$, 
    \begin{align*}
        -\frac{1}{\pi}\emph{\text{Im}}\,\Xi(k) = \frac{\mathrm{i}}{2\pi}(\Xi(k) - \Xi(-k)) = \xi(k).
    \end{align*}
\end{remark}

\begin{remark}\label{Determine the upperbound}
    When applying the formula \eqref{slp and matrix} to compute the Casimir energy, one has to truncate this integral. Therefore, it is necessary to determine a proper upper bound 
    for the Casimir integration. The method for determining the upper bound of the integration is inspired by the asymptotic decay behavior of its integrand function.  
    By Figure \ref{Distinct:The integrand decays exponentially} and Figure \ref{The integrand decays exponentially}, the integrand value $\log\det\mathrm{V}(\mathrm{i}k)\tilde{\mathrm{V}}(\mathrm{i}k)^{-1}$ shares the same trend with $e^{-2Zk}$, this inspires us to apply the function $f(k) = Ce^{-2Zk}$ to fit the curve of the estimated integrand
    values. With the coefficient $C$ determined, one can estimate the absolute error for approximating the Casimir integral by computing:  
    \begin{align} \label{determine ub}
        \epsilon = \int_{\kappa}^{\infty}f(k)dk \approx \frac{Ce^{-2Z\kappa}}{2Z},
    \end{align}
    where $\kappa$ is the upper bound of the integration. Recall that we have changed the variable from $k$ to $y = e^{-k}$ when applying the normal trapezodial rule. This upper bound $\kappa$ corresponds to the lowerbound of $y$.
    
\end{remark}
% \begin{remark}
%     Note that the integral $\frac{\hbar c}{2}\int_{0}^{\infty}\xi(k)dk$ in \eqref{KSSF and CasE} is not Lebesgue convergent and requires regularisation for its numerical evaluation. The right-hand side integral does not suffer from this issue.
% \end{remark}

% {\color{red} Corrected to here}

\subsection{Galerkin discretization and boundary element spaces}
In order to compute the integral \eqref{KSSF and CasE}, we need to compute the log determinant of the operators $V_{k}\tilde{V}_{k}^{-1}$. In this section we discuss Galerkin discretisations to compute this quantity.

Define the 
triangulation $\mathcal{T}_{h}$ of the boundary surface $\Gamma$ with triangular surface elements $\tau_{l}$ and associated nodes $\boldsymbol{x}_{i}$ 
s.t. $\overline{\mathcal{T}_{h}} = \bigcup_{l}\overline{\tau_{l}}$, where $h$ is the mesh size and define the space of the continuous piecewise linear functions
\begin{align*}
    P_{h}^{1}(\Gamma) = \{v_{h}\in C^{0}(\Gamma): v_{h}|_{\tau_{l}}\in\mathbb{P}_{1}(\tau_{l}), \ \text{for} \ \ \tau_{l}\in\mathcal{T}_{h}\},
\end{align*}
where $\mathbb{P}_{1}(\tau_{l})$ denotes the space of polynomials of order less than or equal to 1 on $\tau_{\ell}$. We have

\begin{align*}
    P_{h}^{1}(\Gamma) := \text{span}\{\phi_{j}\} \subset H^{-\frac{1}{2}}(\Gamma)
\end{align*}
with 
\begin{align*}
    \phi_{j}(\boldsymbol{x}_{i}) = \begin{cases}
        1, & i = j,\\
        0, & i\neq j
    \end{cases}
\end{align*}
being the nodal basis functions.

\begin{remark}
Since $H^{-1/2}(\Gamma)$ does not require continuity we could use a space of simple piecewise constant functions. The reason why we choose piecewise linear functions is the size of the arising matrix systems for dense calculations. The computation of the log-determinant requires $\mathcal{O}(n^3)$ operations, where $n$ is the dimension of our approximation basis. For sphere-like and other similar geometries there are in practice roughly twice as many triangles as nodes in the mesh. Hence, while the assembly cost with piecewise linear functions is higher, the resulting matrix has only half the dimension, resulting in roughly a factor eight reduction of computational complexity for the log determinant. A disadvantage is that on geometries with corners or edges the converges close to these singularities is suboptimal with continuous piecewise linear functions.
\end{remark}

Having defined the basis function $\phi_j$, we can represent each element inside the Galerkin discretization form. Assume there are $N$ objects,
then the matrix of the operator $V_k$ is an $N$ by $N$ block matrix, written as 
\begin{align}\label{matrix V}
    \mathsf{V}(k) = \begin{bmatrix}
        \mathsf{V}_{11}(k) & \mathsf{V}_{12}(k) & \cdots & \mathsf{V}_{1N}(k) \\
        \mathsf{V}_{21}(k) & \mathsf{V}_{22}(k) & \cdots & \mathsf{V}_{2N}(k) \\
        \vdots & \vdots & \ddots & \vdots \\
        \mathsf{V}_{N1}(k) & \mathsf{V}_{N2}(k) & \cdots & \mathsf{V}_{NN}(k) \\
\end{bmatrix}
\end{align}
and the matrix $\tilde{V}_{k}$ is the diagonal part of $V_{k}$:
\begin{align}\label{matrix tilde V}
    \tilde{\mathsf{V}}(k) = \begin{bmatrix}
        \mathsf{V}_{11}(k) & 0      & \cdots & 0 \\
    0      & \mathsf{V}_{22}(k) & \cdots & 0\\
    \vdots & \vdots & \ddots & \vdots \\
    0      & 0      & \cdots & \mathsf{V}_{NN}(k) \\
\end{bmatrix}.
\end{align}
Therefore, the element in the $m$th row and $n$th column of the block matrix $\mathsf{V}_{ij}(k)$ is 
\begin{align}\label{Elements in matrix V}
    \mathsf{V}_{ij}^{(m,n)} (k) = \langle V_{ij}(k)\phi_{n}^{(j)}, \phi_{m}^{(i)}\rangle = 
    \int_{\Gamma_{j}}\phi_{m}^{(i)}(\boldsymbol{x})\int_{\Gamma_{i}}g_{k}(\boldsymbol{x}, \boldsymbol{y})\phi_{n}^{(j)}(\boldsymbol{y})dS_{\boldsymbol{y}}dS_{\boldsymbol{x}},
\end{align}
where $\boldsymbol{\phi}^{(i)} = \begin{bmatrix}
    \phi_{1}^{(i)} & \phi_{2}^{(i)} & \dots & \phi_{N}^{(i)}
\end{bmatrix}$ is the set of basis functions defined on the $i$th object and $\langle \cdot, \cdot \rangle$
denotes the standard $L^{2}(\Gamma)$ inner product.

The value of $\Xi(\mathrm{i}k) = \log\det(\mathsf{V}(\mathrm{i}k)\tilde{\mathsf{V}}(\mathrm{i}k)^{-1})$ can now be explicitly computed by evaluating the corresponding log determinants.

The function $\Xi(\mathrm{i}k)$ has a very favourable decay behaviour for growing $k$ that we can use to limit the number of quadrature points necessary to evaluate the corresponding Casimir integral, namely under certain convexity assumptions on the obstacles it holds that
$$
\Xi(\mathrm{i}k) = \mathcal{O}(e^{-2Zk}).
$$
Here, $Z$ is the minimum distance between the obstacles  \cite[Theorem 4.1]{fang2022trace}.

This result can be justified heuristically, using a simple matrix perturbation argument. Consider a symmetric matrix $A$ partitioned as
$$
A = \begin{bmatrix}A_1 & 0\\
                              0   & A_2
       \end{bmatrix}.
$$
and a symmetric matrix $E$ partitioned as
$$
E= \begin{bmatrix}0 & E_1^T\\
     E_1 & 0
     \end{bmatrix}
$$
Then it holds for the $i$th eigenvalue $\lambda_i(A)$ and the $i$th eigenvalue $\lambda_i(A+E)$ that
$$
|\lambda_i(A) - \lambda_i(A+E)| \leq \frac{\|E\|^2}{\text{gap}_i},
$$
where $\text{gap}_i$ is the distance of $\lambda_i(A)$ to the spectrum of $A_2$ if $\lambda_i(A)$ is an eigenvalue of $A_1$, and to the spectrum of $A_1$ if $\lambda_i(A)$ is an eigenvalue of $A_2$. Details can be found in  
\cite{mathias1998quadratic}.
% {\color{red} The reference to Roy Matthias is suddenly missing}.

\begin{figure}[H]
    \centering
    \includegraphics[scale = 0.7]{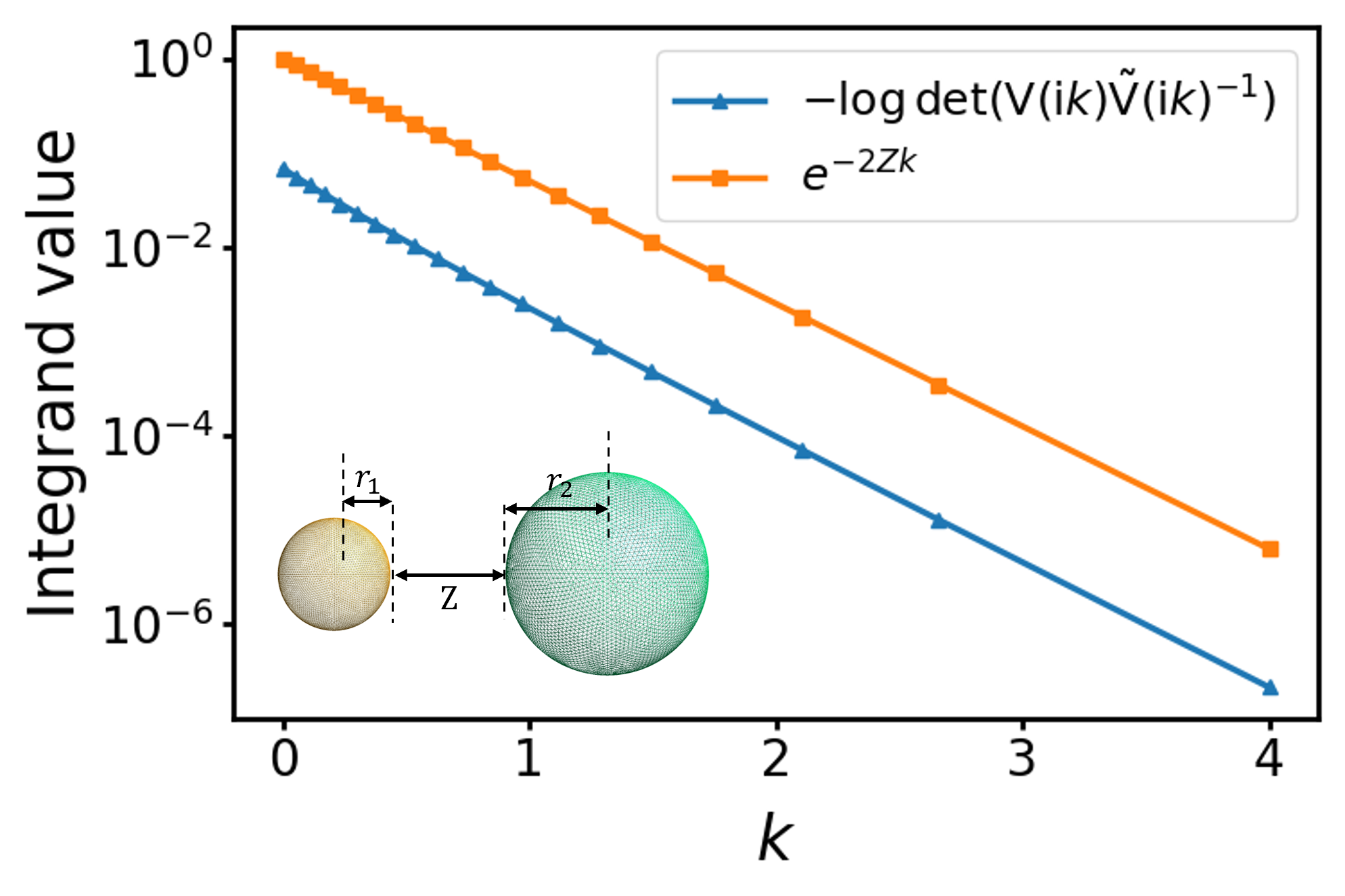}
    \caption{Exponential decay of $\Xi(\textrm{i}k)$ for two distinct spheres with radii $r_{1} = 0.5$ and $r_{2} = 1$ and minimum distance $Z=1.5$. The red line is the decay bound and the blue line is the actual decay.}
    \label{Distinct:The integrand decays exponentially}
\end{figure}

Now assume that we have two different obstacles. Then we have $A_1 = \mathsf{V}_{11}(\mathrm{i}k)$, $A_2 = \mathsf{V}_{22}(\mathrm{i}k)$ and $E_1 = \mathsf{V}_{21}(\mathrm{i}k)$ as the matrix of cross interactions 
between the two obstacles. For complex wavenumbers $\mathrm{i}k$, the Green's function between two obstacles decays exponentially like $e^{-Zk}$, where $Z$ is the minimal distance between them, 
resulting in a matrix perturbation result of the form $|\lambda_i(\mathsf{V}) - \lambda_i(\tilde{\mathsf{V}})| = O(e^{-2Zk})$ for increasing $k$ (see Figure \ref{Distinct:The integrand decays exponentially}
), from which the corresponding perturbation result for the log determinant follows.

This purely linear algebraic consideration is not fully robust as it ignores the importance of the eigenvalue gap in the perturbation result. But we can 
heuristically explain the $\text{gap}$ as follows. On the continuous level the perturbations $E_1$ and $E_2$ are compact, so the tail end of the spectrum 
that converges to zero with small values of $\text{gap}_i$, is little affected by $E$, and the corresponding eigenvalues have a contribution of 
$\log \left|\frac{\lambda_i(A)}{\lambda_i(A+E)}\right| \approx 0$ to the value of $\Xi$. The relevant eigenvalues are the larger ones who for distinct obstacles 
have a sufficiently large value of $\text{gap}_i$.

\begin{figure}[H]
    \centering
    \includegraphics[width = \textwidth]{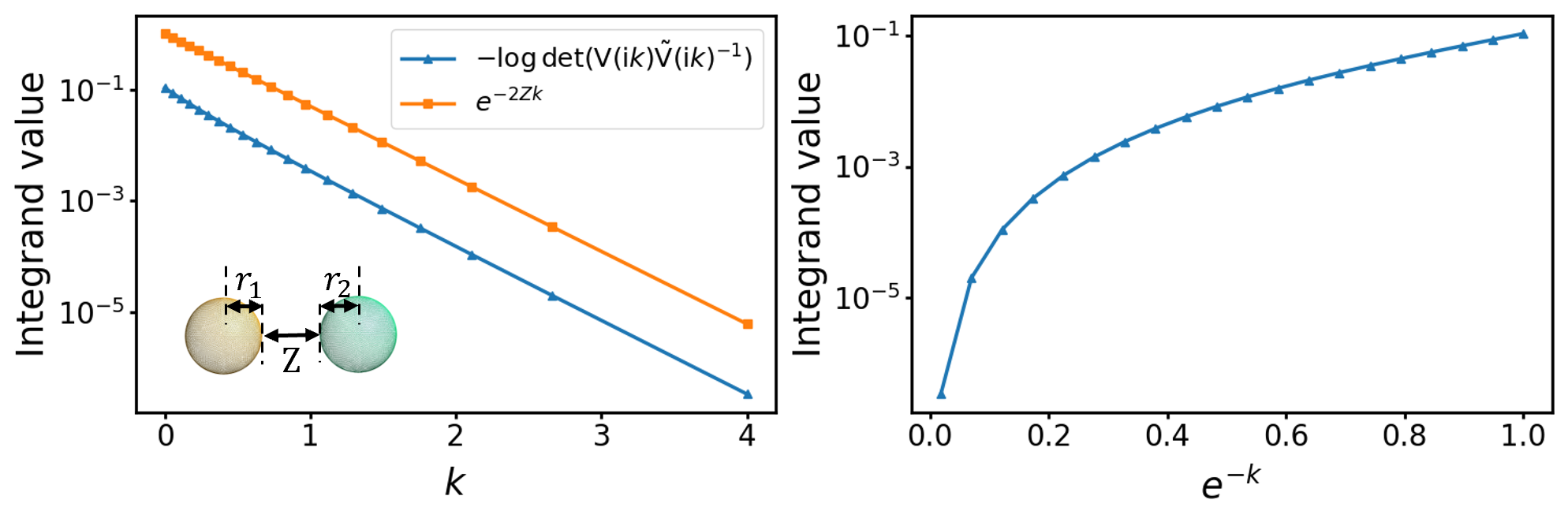}
    \caption{(Left) Exponential decay of $\Xi(\textrm{i}k)$ for two identical spheres with radius $r_1 = r_2 =1$ and minimum distance $Z=1.5$. The red line is the decay bound and the blue line is the actual decay. (Right) The integrand $\Xi(ik)$ after varlable transformation to apply a numerical trapezoid rule for its evaluation.}
    \label{The integrand decays exponentially}
\end{figure}

While the linear algebra argument is useful to give a heuristical explanation, it is not as rigorous as the analytical result in \cite{fang2022trace}. In particular, we want to emphasize that the exponential decay bound with the quadratic factor also holds if the two obstacles are identical, which is not obvious from pure linear algebraic considerations. An example of this is given in Figure \ref{The integrand decays exponentially}.

The exponentially decay property motivates a simple change of variables through $y = e^{-k}$ in the integrant $\Xi(\mathrm{i}k) = \log\det(\mathsf{V}(\mathrm{i}k)\tilde{\mathsf{V}}(\mathrm{i}k)^{-1})$, 
which after transformation we can numerically evaluate with a simple trapezoidal rule. Figure \ref{The integrand decays exponentially} (Right) plots the integrand with regard to the new variable $y$.

\section{Efficient methods for computing $\log\det(\mathsf{V}(\mathrm{i}k)\tilde{\mathsf{V}}(\mathrm{i}k)^{-1})$}\label{Krylov subspace for generalized eigenvalue problem}
% !TEX root =  main.tex
By Section \ref{Numerical methods for computing the Casimir energy}, to compute the Casimir energy, it is necessary to evaluate the term
$\log\det(\mathsf{V}(\mathrm{i}k)\tilde{\mathsf{V}}(\mathrm{i}k)^{-1})$ 
with different values of $k$. In this section, several efficient methods will be introduced to compute this log determinant.

\begin{figure}[H]
    \centering
    \includegraphics[scale = 0.5]{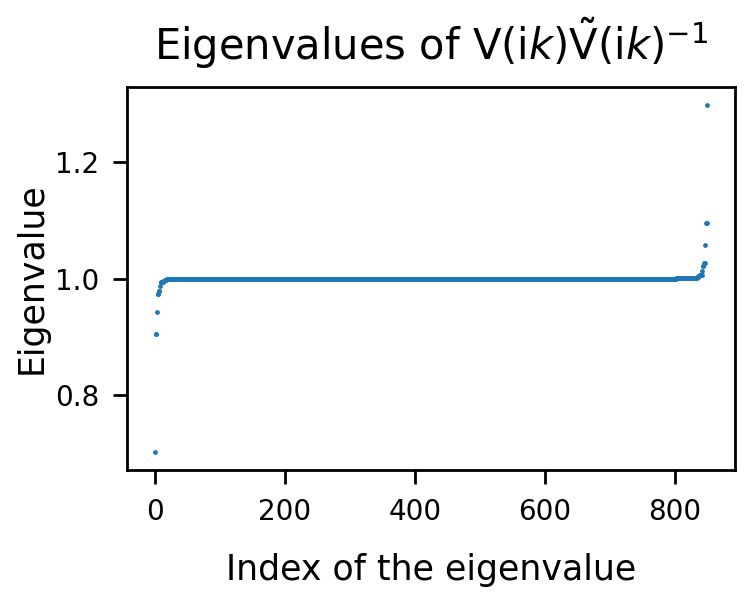}
    \caption{The eigenvalues of the matrix $\mathsf{V}(\mathrm{i}k)\tilde{\mathsf{V}}(\mathrm{i}k)^{-1}$ when $\mathrm{i}k = 0.8\mathrm{i}$.
    The scatterers are two spheres with equal radii $r_{1} = r_{2} = 1$ and the minimal distance between them is $Z = 0.5$. The grid size of the mesh is $h = 0.2$.}
    \label{eigenvalues of VVtilde}
\end{figure}

The log determinant of the matrix $\mathsf{V}(\mathrm{i}k)\tilde{\mathsf{V}}(\mathrm{i}k)^{-1}$ is equal to the sum of the logarithm of the eigenvalues of 
$\mathsf{V}(\mathrm{i}k)\tilde{\mathsf{V}}(\mathrm{i}k)^{-1}$. Since $\tilde{\mathsf{V}}(\mathrm{i}k)$ is a compact perturbation of $\mathsf{V}(\mathrm{i}k)$,
most of the eigenvalues of the matrix $\mathsf{V}(\mathrm{i}k)\tilde{\mathsf{V}}(\mathrm{i}k)^{-1}$ are close to 1 
(shown in Figure \ref{eigenvalues of VVtilde}) and contribute little to the value of the Casimir energy. Hence, we do not need to compute all eigenvalues but only
the extremal ones, making subspace methods such as Krylov solvers attractive for this problem.

In what follows we demonstrate and compare iterative solver approaches based on standard Arnoldi iterations \cite{arnoldi1951principle, MR3396212}, and based on the 
inverse free Krylov subspace method \cite{golub2002inverse, money2005algorithm}.
We will also discuss on acceleration strategy which is based on the idea of recycling projection bases from one quadrature point to the next. 
\subsection{Method I: Standard Arnoldi method}
The first efficient method for solving our eigenvalue problem $\mathsf{V}(\mathrm{i}k)\tilde{\mathsf{V}}(\mathrm{i}k)^{-1}\boldsymbol{x} = \lambda\boldsymbol{x}$ 
is the Arnoldi method \cite[Section 6.2]{MR3396212}. The idea of this method is to use Arnoldi iterations to construct the Krylov subspace  
$K_{m}(\mathsf{V}(\mathrm{i}k)\tilde{\mathsf{V}}(\mathrm{i}k)^{-1}, \boldsymbol{b})$, where $\boldsymbol{b}$ is some 
initial vector and $m$ is the dimension of this Krylov subspace, and to then compute the eigenvalues of the resulting projected Hessenberg matrix $H_m$ (see \cite{saad2011numerical}).
These eigenvalues have a good approximation on the extreme eigenvalues of $\mathsf{V}(\mathrm{i}k)\tilde{\mathsf{V}}(\mathrm{i}k)^{-1}$ \cite[Proposition 6.10, Theorem 6.8]{MR3396212}. 

The main cost of this standard Arnoldi method is the computation of the Krylov subspace $K_{m}$. In this process, one has to compute the matrix-vector product 
$\tilde{\mathsf{V}}^{-1}\boldsymbol{y}$, for some vector $\boldsymbol{y}$, which is equivalent to solve the linear system $\tilde{\mathsf{V}}\boldsymbol{x} = \boldsymbol{y}$. This can be 
efficiently implemented as the matrix $\tilde{\mathsf{V}}$ is a block diagonal matrix. Therefore, we just need to compute the LU decomposition for each diagonal 
block, $\mathsf{V}_{jj}$ rather than the whole system matrix and apply the forward and backward substitution to solve the linear system 
$\mathsf{V}_{jj}\,\boldsymbol{x}_j = \boldsymbol{y}_j$. Note that if all the scatterers are identical, one only needs to compute one diagonal block and one LU decomposition. 

\subsection{Method II: Inverse-free Krylov subspace method}
An alternative to the standard Arnoldi method is the inverse-free projection method, which is also based on the Arnoldi iterations but without computing any 
matrix inversions. Consider the eigenvalue problem $\mathsf{V}(\mathrm{i}k)\tilde{\mathsf{V}}(\mathrm{i}k)^{-1}\boldsymbol{x} = \lambda\boldsymbol{x}$,
it is equivalent to the following generalized eigenvalue problem:
\begin{align}\label{GEP}
    \mathsf{V}(\mathrm{i}k)\tilde{\boldsymbol{x}} = \lambda \tilde{\mathsf{V}}(\mathrm{i}k)\tilde{\boldsymbol{x}}.
\end{align}

An important property of this problem is that as we are only interested in $\mathrm{i}k$ along the imaginary axis, the corresponding matrix $\tilde{\mathrm{V}}(\mathrm{i}k)$ is positive definite,
and $\mathrm{V}(\mathrm{i}k)$ is still symmetric.

In \cite{golub2002inverse, money2005algorithm},
the authors proposed an inverse-free Krylov subspace method for computing a few extreme eigenvalues of the symmetric definite generalized eigenvalue problem.
The following algorithm summarizes the method.

\begin{algorithm}[H]
    \SetAlgoLined
    Input: Symmetric matrix $A\in\mathbb{R}^{n\times n}$, s.p.d matrix $B\in\mathbb{R}^{n\times n}$, an initial approximation $\boldsymbol{x}$ with $||\boldsymbol{x}|| = 1$,
    a given shift $\rho$ and the dimension of the Krylov subspace $m\geq 1$\\
    Output: A set of approximate eigenvalues of $A\boldsymbol{x} = \lambda B\boldsymbol{x}$ and associated eigenvectors.\\
    \begin{algorithmic}[1]
        
        \STATE Construct a basis $Z_{m}$ for the Krylov subspace $K_{m} = \text{span}(\boldsymbol{x}, (A - \rho B)\boldsymbol{x}, \dots, (A - \rho B)^{m-1}\boldsymbol{x})$ with dimension $m$
        \STATE Project $A$ and $B$ on $Z$: $A_{m} = Z_{m}^{T}(A - \rho B)Z_{m}$, $B_{m} = Z_{m}^{T}BZ_{m}$
        \STATE Compute all the eigenpairs $\{(\tilde{\lambda}_{i}, \boldsymbol{x}_{i})\}_{i = 1, \dots, m}$ for the matrix pencil $(A_{m}, B_{m})$
        \STATE Reverse the shift to obtain $\lambda_{i} = \tilde{\lambda}_{i} + \rho$.
        \end{algorithmic}
    \caption{Inverse-free Krylov subspace method for computing multiple extreme eigenvalues of the generalized eigenvalue problem $A\boldsymbol{x} = \lambda B\boldsymbol{x}$}
    \label{Alg for computing the evals kry}
    \end{algorithm}

% {\color{red} Why do you say extreme? Isn't the algorithm finding eigenvalues close to the shift?} 
Algorithm \ref{Alg for computing the evals kry} approximates $m$ eigenvalues close to the shift $\rho$ 
for the matrix pencil $(A,B)$, where $m$ is the dimension of the 
Krylov subspace $K_{m}$ in Step 1, Algorithm \ref{Alg for computing the evals kry}. The question is what shift strategy to use for $\rho$. In numerical experiments it turned
out that for the KSSF problem choosing $\rho=1$ sufficiently approximate the eigenvalues that have the main contribution to $\log\det(\mathsf{V}(\mathrm{i}k_{j})\tilde{\mathsf{V}}(\mathrm{i}k_{j})^{-1}) $.
Additionally, the main cost of this inverse free Krylov subspace method is the computation of the Krylov subspace and the projection of the matrices $A$ and $B$. In our case these are large dense matrices representing
integral operators. 
% The dominant cost of the inverse-free Krylov subspace method is that of the involved matrix-vector products with the integral operators in the process of the
% computation of the Krylov subspace $K_m$ and the projection of the shifted matrix $(A−\rho B)$ and $B$ onto the orthogonal basis of $K_m$.

\subsection{Recycling Krylov subspace based variant}

The main cost of the standard Arnoldi method and inverse-free method comes from the matrix-vector products (matvecs) in the Arnoldi iterations, where the 
involving matrices are large and dense as they represent discretized integral operators. In order to reduce the computational cost of a Krylov subspace basis 
for each wavenumber $\mathrm{i}k$, we introduce a subspace recycling based method for speeding up the computational process. This can be regarded 
as a variant of these two methods.

This recycling strategy is based on the idea that a Krylov subspace for a previous quadrature
point in the KSSF integral will be a good approximation to a Krylov subspace for the current quadrature point. We initially compute a Krylov basis for the wavenumber $\mathrm{i}k_{1}$ associated with the first
quadrature point. We then extract several eigenvectors associated with the extremal eigenvalues based on Algorithm \ref{Alg for computing the evals kry} and then orthogonalize to obtain an initial approximation
basis for the wavenumber $\mathrm{i}k_2$. For this wavenumber we project the matrices onto the recycled basis, compute approximate eigenpairs $(\tilde{\lambda}_i, \tilde{\mathbf{x}}_i)$ and then extend the subspace
basis with the residuals $\boldsymbol{r}_i = \mathsf{V}(\mathrm{i}k)\tilde{\boldsymbol{x}}_i - \tilde{\lambda}_i\tilde{\mathsf{V}}(\mathrm{i}k)\tilde{\boldsymbol{x}}_i$. With the extended subspace we recompute the eigenpairs for the second wavenumber's
case and extract eigenvectors as starting basis for the third wavenumber, and so on.

\subsection{Comparison of efficient methods for computing $\log\det(\mathsf{V}(\mathrm{i}k)\tilde{\mathsf{V}}(\mathrm{i}k)^{-1})$}
In this section, we compare the performance of standard Arnoldi and inverse-free Krylov subspace method and their recycled variants on computing the log 
determinant term of $\mathsf{V}(\mathrm{i}k)\tilde{\mathsf{V}}(\mathrm{i}k)^{-1}$. As the dominant cost of these methods is the matrix-vector products (matvec) 
associated with the discretized boundary integral operators, 
we will also compare the number of the matvecs between these methods. All the tests are performed on two spheres with equal radii $r_1 = r_2 = 1$. 
The sphere meshes are refined with size $h = 0.1$ and this results in the matrix size $\mathrm{V}(\mathrm{i}k)$ being $3192\times 3192$. Again, the minimum distance between them is denoted by $Z$, which is set as 0.5, 1.5 and 3.0. 
The number of the quadrature points is 20.

\begin{figure}[H]
    \centering
    \includegraphics[width = \textwidth]{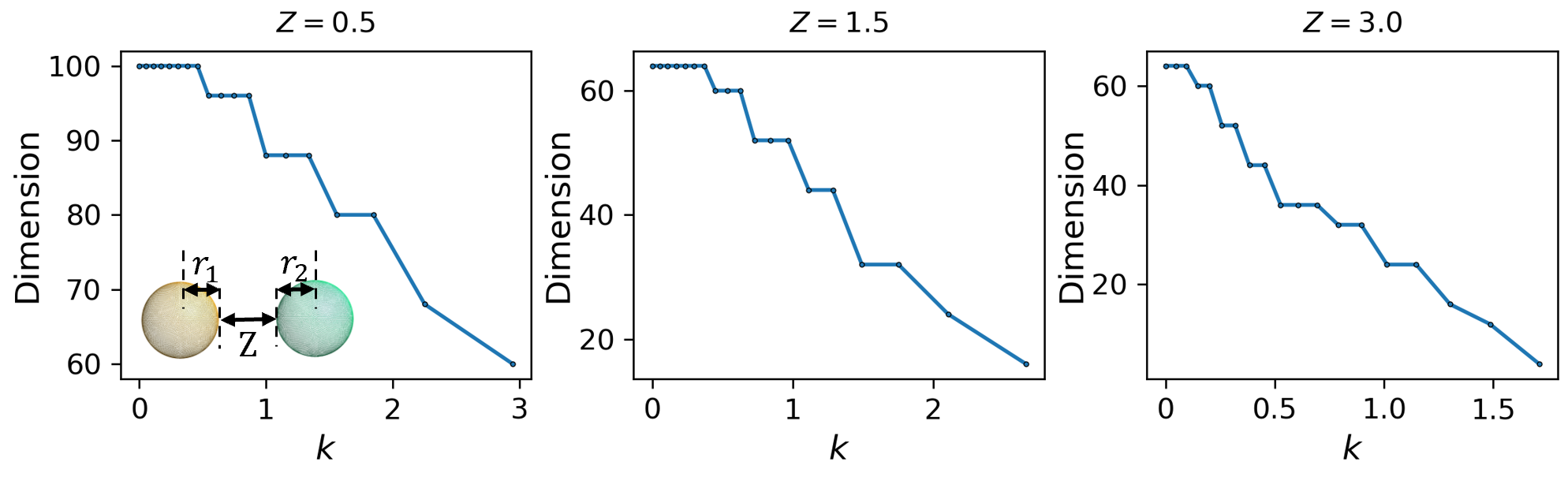}
    \caption[Caption for LOF]{The dimension of the extended subspace in the inverse-free Krylov subspace\protect\footnotemark with subspace recycled for each $k$ in $\log\det(\mathsf{V}(\mathrm{i}k)\tilde{\mathsf{V}}(\mathrm{i}k)^{-1})$, which is equal to the number of the recycled 
    eigenvectors plus the number of the residuals $\{\mathbf{r}_{i}\}_{i}$. The recycled eigenvector has the corresponding eigenvalue whose logarithm
    is larger than $10^{-s}$, where $s = 3, 4, 5$ when $Z = 0.5, 1.5, 3.0$, respectively.}
    \label{Dimension_of_the_extended_subspace}
\end{figure}

We start by comparing the relative error for approximating the $\log\det\mathsf{V}(\mathrm{i}k)\tilde{\mathsf{V}}(\mathrm{i}k)^{-1}$ using all these methods. 
 The dimension of the Krylov subspace $K_m$ in all the algorithms is set to $m = 100$ to ensure that the values computed by these efficient methods maintain a precision of at least three significant 
digits, aligning closely with those obtained through direct dense computation.
For the methods with subspace recycled, the number of the recycled eigenvectors is not fixed but depends on the number of the relevant eigenvalues in each wavenumber case. 
In our experiments, we only recycle the eigenvector whose corresponding eigenvalue has the logarithm value greater than $10^{-s}$, where $s = 3, 4, 5$ when $Z = 0.5$, 1.5, 3.0, respectively, 
in which case the estimates of the log determinant have at least three significant digits 
match with the ones obtained from the direct computations. In this direct method, the determinant of the matrix is densely computed using LU decomposition. With these settings, the number of the recycled
eigenvectors becomes less and less as the $k$ gets larger.  Figure \ref{Dimension_of_the_extended_subspace} plots the dimension of the extended subspace
for each wavenumber case. It is equal to the number of the recycled eigenvectors plus the number of the residuals $\{\mathbf{r}_{i}\}_{i}$.

\footnotetext{Same figure also applies for standard Arnoldi methods.}

Table \ref{Table lists the logdet} lists the relative error for approximating the value of $\log\det(\mathsf{V}(\mathrm{i}k)\tilde{\mathsf{V}}(\mathrm{i}k)^{-1})$ 
computed via the inverse-free Krylov subspace method and standard Arnoldi method with or without recycling the subspace. The reference value is computed by the 
direct dense computation of the log determinant. The wavenumbers $\mathrm{i}k$ are chosen to be associated with the first five consecutive quadrature points, 
whose corresponding log determinant values account for a great proportion in the Casimir integral. 

This table indicates that with the settings above, one can have at least three significant digits accuracy and the accuracy of the methods with subspace 
recycled is similar to the ones without any recycling processes. As for the performance of these methods and their variants at larger quadrature points, we 
cannot always have three digits accuracy. However, this will not affect the estimates of the Casimir energy too 
much as their corresponding log determinant value is relatively smaller than the others and contributes very little to the Casimir energy. 
 
\begin{table}[H]
    \centering
    \begin{tabular}{ |M{1.5cm}|M{2.0cm}|M{2.2cm} |M{2.2cm}|M{3cm}|M{3cm}|M{2.2cm}| } 
    \hline
    Distance $Z$ & Wavenumber $k$ &  Inverse-free (no recycling) & Inverse-free (recycling) & Standard Arnoldi (no recycling) & Standard Arnoldi (recycling)\\
    \hline
    \multirow{5}{4em}{$Z = 0.5$}   & 0        & $9.79\times 10^{-4}$  & $9.79\times 10^{-4}$  &$9.29\times 10^{-4}$ &$9.29\times 10^{-4}$\\ 
                                   & 0.0540   & $9.67\times 10^{-4}$  & $9.78\times 10^{-5}$  &$4.91\times 10^{-5}$ &$1.37\times 10^{-6}$\\ 
                                   & 0.111    & $1.22\times 10^{-3}$  & $2.79\times 10^{-5}$  &$5.29\times 10^{-5}$ &$5.17\times 10^{-6}$\\ 
                                   & 0.171    & $1.15\times 10^{-3}$  & $2.42\times 10^{-5}$  &$2.78\times 10^{-5}$ &$8.45\times 10^{-5}$\\ 
                                   & 0.236    & $1.25\times 10^{-3}$  & $9.10\times 10^{-6}$  &$1.12\times 10^{-4}$ &$2.76\times 10^{-5}$\\ 
    \hline
    \hline
    \multirow{5}{4em}{$Z = 1.5$}   & 0        & $9.48\times 10^{-4}$  & $9.54\times 10^{-4}$  &$3.41\times 10^{-7}$ &$3.41\times 10^{-7}$\\ 
                                   & 0.0530   & $1.02\times 10^{-3}$  & $2.87\times 10^{-4}$  &$5.89\times 10^{-7}$ &$3.97\times 10^{-8}$\\ 
                                   & 0.109    & $1.16\times 10^{-3}$  & $1.80\times 10^{-4}$  &$1.45\times 10^{-8}$ &$2.35\times 10^{-4}$\\ 
                                   & 0.168    & $1.25\times 10^{-3}$  & $1.35\times 10^{-4}$  &$2.70\times 10^{-6}$ &$1.06\times 10^{-4}$\\ 
                                   & 0.231    & $1.33\times 10^{-3}$  & $4.77\times 10^{-5}$  &$3.14\times 10^{-7}$ &$4.87\times 10^{-5}$\\ 
    \hline
    \hline
    \multirow{5}{4em}{$Z = 3.0$}   & 0        & $1.38\times 10^{-3}$  & $1.38\times 10^{-3}$  &$8.55\times 10^{-12}$ &$8.55\times 10^{-12}$\\ 
                                   & 0.0465   & $1.54\times 10^{-3}$  & $4.34\times 10^{-4}$  &$3.46\times 10^{-9}$  &$2.61\times 10^{-5}$\\ 
                                   & 0.0954   & $1.81\times 10^{-3}$  & $2.89\times 10^{-4}$  &$5.02\times 10^{-10}$ &$5.43\times 10^{-7}$\\ 
                                   & 0.146    & $2.13\times 10^{-3}$  & $2.35\times 10^{-4}$  &$4.82\times 10^{-8}$  &$2.50\times 10^{-5}$\\ 
                                   & 0.200    & $2.54\times 10^{-3}$  & $2.13\times 10^{-4}$  &$5.07\times 10^{-9}$  &$1.59\times 10^{-5}$\\ 
    \hline
    \end{tabular}
    \caption{Relative error for approximating the value of $\log\det(\mathrm{V}(\mathrm{i}k)\tilde{\mathrm{V}}(\mathrm{i}k)^{-1})$ on the wavenumbers associated with the first five consecutive 
    quadrature points via the inverse-free Krylov subspace and standard Arnoldi methods with/without subspace recycled. The shift is set as $\rho = 1$ for the inverse-free method. The recycled eigenvector has the corresponding eigenvalue whose logarithm
    is larger than $10^{-s}$, where $s = 3, 4, 5$ when $Z = 0.5, 1.5, 3.0$, respectively.}
    \label{Table lists the logdet}
    \end{table}
    
    Recall that the main cost in these algorithms is from the computation of the Krylov basis and the matrix projections. For large problems, 
    the dominating cost is the involved matrix-vector products with the discretized integral operators. We count the number of matvecs associated with the 
    discretized integral operators ($\mathsf{V}_{ij}$) for each algorithm and the results are summarized in Table \ref{4methods_matvecs}.

    \begin{table}[H]
        \centering
    
    \begin{tabular}{ |P{3cm}|P{3.8cm}|P{3cm}|P{3.6cm}|}
    
        \hline
        \multicolumn{2}{|c|}{Inverse-free Krylov subspace method}& \multicolumn{2}{c|}{Standard Arnoldi method} \\
        \hline
      Without recycling &  With recycling & Without recycling& With recycling\\
        \hline
        $(6m - 2)N_q$  & $(6m - 2) + 12\sum\limits_{i = 1}^{N_q-1}s_{i}$   & $(4m - 4)N_q$ &   $(4m - 4) + 8\sum\limits_{i = 1}^{N_q-1}s_{i}$ \\
        \hline
      \end{tabular}
      \bigskip
      \caption{The number of matvecs associated with the discretized integral operators inside the inverse-free Krylov subspace and standard Arnoldi methods with or without recycling subspace.
      $N_q$ is the number of wavenumbers/quadrature points, $m$ is the dimension of the Krylov subspace for the first wavenumber (in recycling case); for all the wavenumbers (in non-recycling case),
      and $s_{i}$ is the number of the recycled eigenvectors 
      from the $i$th wavenumber's case (in recycling case)}
      \label{4methods_matvecs}
    \end{table}

\begin{figure}[H]
        \centering
        \includegraphics[width = \textwidth]{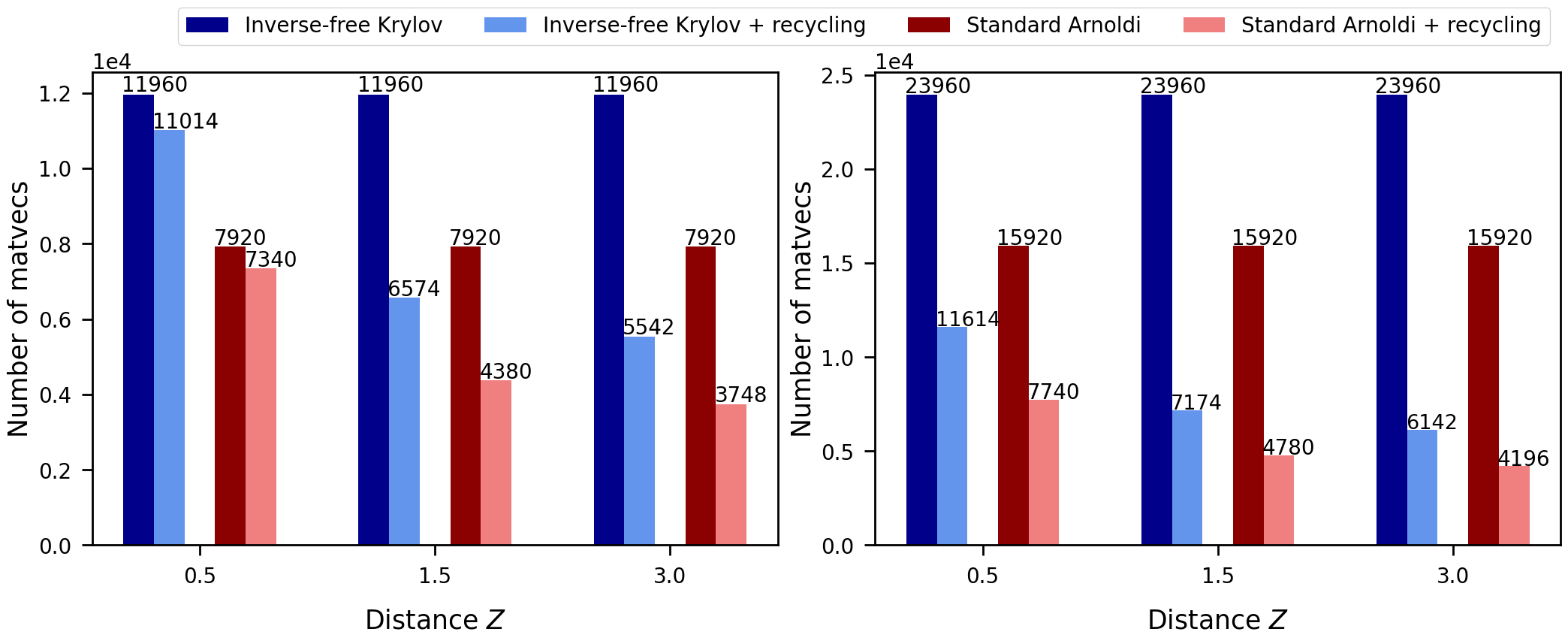}
        \caption{The number of matvecs inside the inverse-free and standard Arnoldi methods with or without recycling subspace when 
        computing the normalized Casimir energy between two spheres with equal radii $R = 1$ and the distance $Z$ is 0.5, 1.5 and 3.0. 
        The number of the quadrature point is $N_q = 20$. The dimension of the Krylov subspace is set as $m = 100$ ({\color{gray} Left}) and 
        $200$ ({\color{gray} Right}).}
        \label{fig:Num_matvec}
    \end{figure}

    In Figure \ref{fig:Num_matvec}, we plot the number of actual matvecs associated with the discretized matrix form $\mathsf{V}_{ij}$, for $i, j = 1, 2, \cdots, N$ in each individual algorithm when computing the Casimir energy between 2 spheres with different different distance $Z$. It shows that the recycling strategy significantly reduces the overall number of matvecs. Although the number of matvecs in standard Arnoldi method with subspace recycled (light red in Figure \ref{fig:Num_matvec}) is smaller than the one of the inverse-free method  with subspace recycled (light blue in Figure \ref{fig:Num_matvec}), one has to compute the LU decomposition for each diagonal block in each Arnoldi iteration, which has cubic complexity. 

%Note that for the recycled methods Algorithm \ref{Alg for computing the evals kry recycled}-\ref{Alg for computing the evals arno recycled}
%we apply different rules for extracting the eigenvectors. For  
%is different which depends on the 

%For the standard Arnoldi method,  it can be noticed that the number of FLOP is cubicly increasing with the size of the matrix 
%no matter the subspace is recycled or not. For the inverse-free Krylov subspace methods, as the wavenumber $k$ increases, the number of the extreme eigenvalues
%decreases which makes the number of extracted eigenvectors decreases as well. Therefore, for the large-scale problems, the inverse-free Krylov subspace method with 
%subspaces recycled would be applied to compute the Casimir energy with lower complexity and desired accuracy.

\section{Numerical experiments for computing Casimir energy}\label{Numerical experiments}
% !TEX root =  main.tex

In this section, we present numerical results illustrating the computation of Casimir energy between two conducting objects. 
The considered objects include spheres, sphere-torus configurations, Menger sponges, ice crystals, and ellipsoids. Various methods are employed for computing 
the Casimir energy, including the inverse-free Krylov subspace, standard Arnoldi methods, and their recycled invariants, along with the Richardson extrapolation 
method. To establish a reference value for the Casimir energy, we utilize the Richardson extrapolation method, commonly employed to obtain higher-order 
estimates at zero grid spacing.
% Reference values for the Casimir energy in each case is computed by grid refinement plus Richardson extrapolation.

%The reference value of the Casimir energy is computed by Richardson extrapolation method which is often used 
%for obtaining the higher-order estimate at zero grid spacing. Denote $\mathcal{E}_{\text{fine}}$ and $\mathcal{E}_{\text{coarse}}$ as the Casimir energy 
%numerically computed from the formula \eqref{KSSF and CasE} by setting the grid size $h$ as $h_{\text{fine}}$ and $h_{\text{coarse}}$ 
%($h_{\text{fine}}<h_{\text{coarse}}$), separately. Then, the high-accuracy result $\mathcal{E}_{\text{extrapolation}}$ can be generated from the following formula:
%\begin{align}\label{Richardson extrapolation}
%    \mathcal{E}_{\text{extrapolation}} \approx  \frac{h_{\text{coarse}}^{2}\mathcal{E}_{\text{fine}} - h_{\text{fine}}^{2}\mathcal{E}_{\text{coarse}}}{h_{\text{coarse}}^{2} - h_{\text{fine}}^{2}}.
%\end{align}

In the case of spheres we also compare with known asymptotic expansions \cite{emig2008casimir}.

\subsection{Sphere-sphere and sphere-torus case}
\begin{figure}[H]
    \hspace*{3cm}\includegraphics[scale = 0.6]{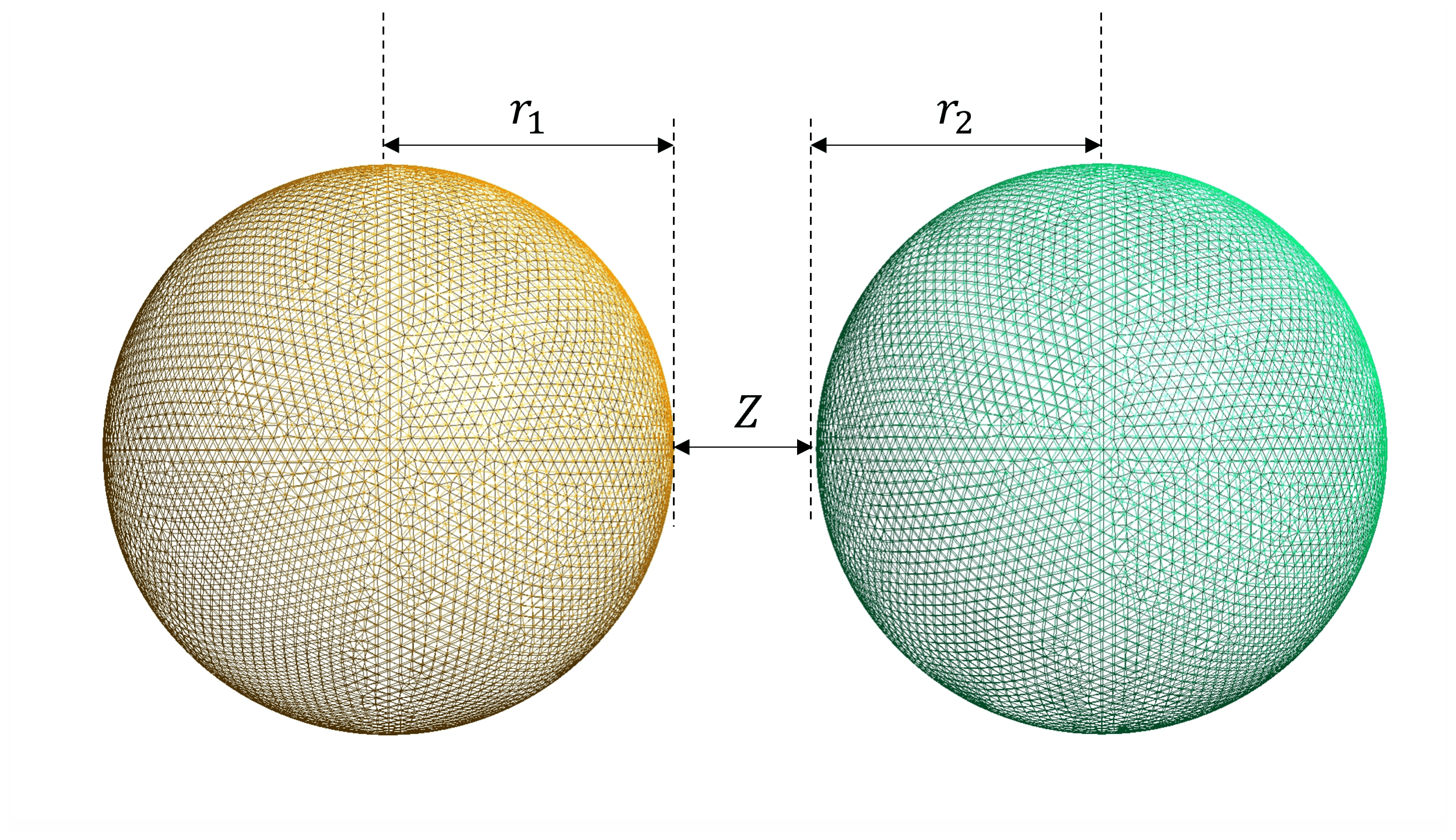}
    \caption{Two spheres with equal radii $r_{1} = r_{2} = 1$ and $Z$ is the minimal distance between them.
    \\ \hspace*{1.2cm} $h_{\text{coarse}} = 0.1$: $\text{dim}(\mathrm{V}(\mathrm{i}k)) = 3192$,  N\textsuperscript{\underline{o}} of elements on both grids $ = 6384$;\\
    \hspace*{1.2cm}$h_{\text{fine}} = 0.05$: $\text{dim}(\mathrm{V}(\mathrm{i}k)) = 12603$,  N\textsuperscript{\underline{o}} of elements on both grids $ = 25180$}
    \label{Two spheres with equal radii}
\end{figure}

Consider two spheres with the equal radii $r_1 = r_2 = 1$ and the spacing $Z$ (see Figure \ref{Two spheres with equal radii}) as the scatterers.
We denote $\Xi_{h}$, the value of $\Xi$ computed under the refinement level with mesh size $h$ and denote $\Xi_{h = 0}$, the higher-order estimate of 
$\Xi_{h}$ at zero grid space, which is computed by Richardson extrapolation,
\begin{align}\label{Extrapolation formula}
    \Xi_{h = 0} \approx \frac{h_{\text{coarse}}^{2}\Xi_{h_\text{fine}}  - h_{\text{fine}}^{2}\Xi_{h_\text{coarse}}}{h_{\text{coarse}}^{2}  - h_{\text{fine}}^{2}},
\end{align}
where $h_{\text{fine}}$ and $h_{\text{coarse}}$ are two different mesh sizes with $h_{\text{fine}} < h_{\text{coarse}}$. In this example, we set $h_{\text{coarse}} = 0.1$ and $h_{\text{fine}} = 0.05$.

\begin{figure}[H]
    \centering
    \includegraphics[width = \textwidth]{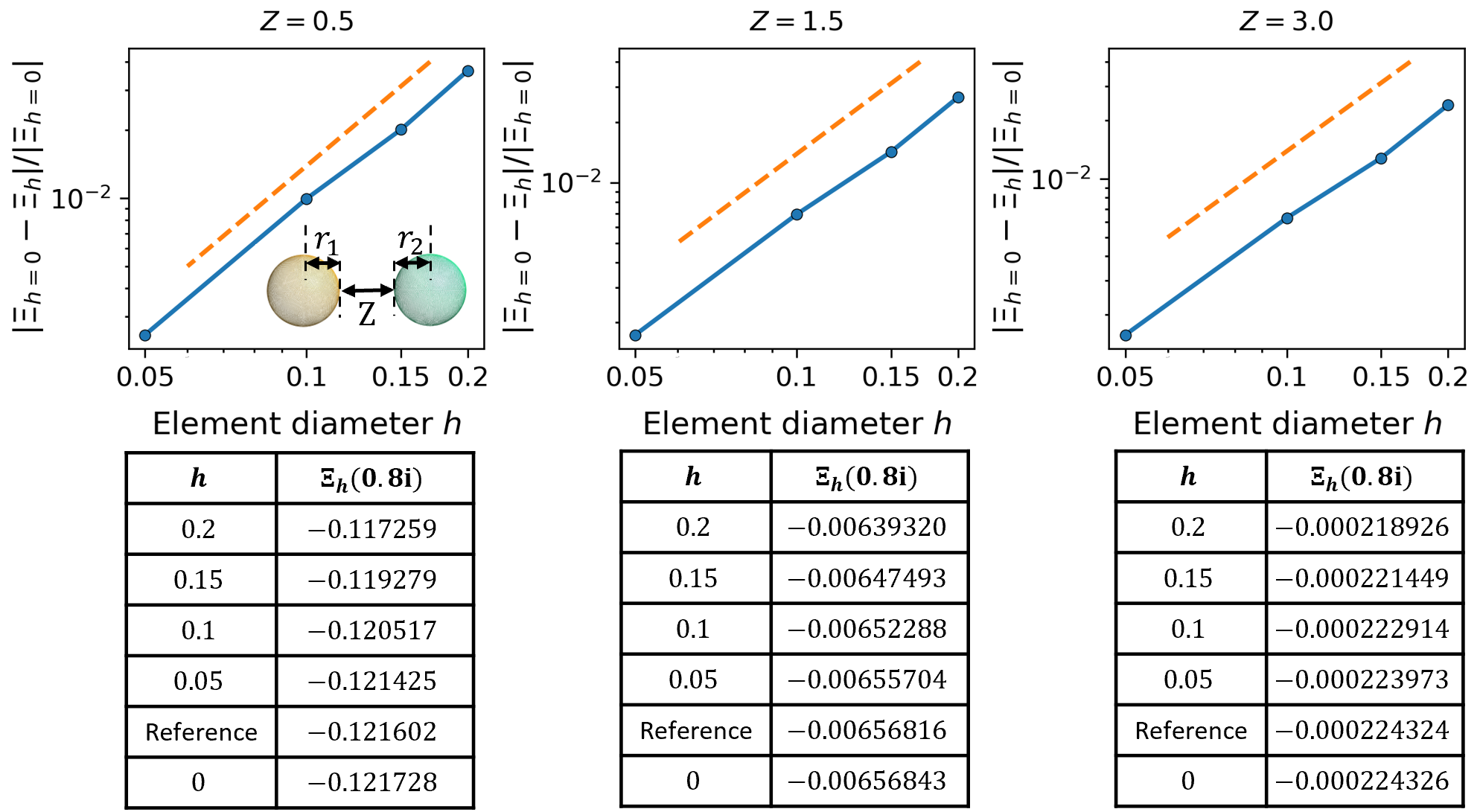}
    \caption{$h$-convergence of $\Xi_{h}(\mathrm{i}k)$ to the extrapolation value $\Xi_{h = 0}(\mathrm{i}k)$ when $\mathrm{i}k = 0.8\mathrm{i}$. The provided reference values are accurate within $0.05\%$. The scatterers are two spheres with equal radii 1 and the distance between them is set as $Z = 0.5$, 1.5 and 3.0.
    The relative distance between $\Xi_{h}$ and $\Xi_{h = 0}$ decreases as we refine the mesh. The dashed line shows order 2 convergence. The tables list the values of $\Xi_{h}(0.8\mathrm{i})$ for $h = 0$, 0.05, 0.1, 0.15 and $0.2$, and the provided reference values.}
    \label{Scalar_Xi_h_conv}
\end{figure}

We begin with validating the construction of the integrand function $\Xi$ of the Casimir integral \eqref{KSSF and CasE} by comparing the value of 
$\Xi_{h}(\mathrm{i}k)$ for different refinement levels with the extrapolation value $\Xi_{h = 0}$ for $\mathrm{i}k = 0.8\mathrm{i}$ (see Figure \ref{Scalar_Xi_h_conv}). In the tables of 
Figure \ref{Scalar_Xi_h_conv}, we also provide reference values computed by discretizing the single-layer boundary integral operators in terms of the spherical harmonic functions as suggested by \cite{kenneth2008casimir}.
They are believed to be accurate within $0.05\%$.

In Figure \ref{Scalar_Xi_h_conv}, one can see that $\Xi_{h}(\mathrm{i}k)$ converges to $\Xi_{h = 0}$ as $h$ increases.
This figure is plotted in log-log scale plot and the slope of these lines is around 2. This numerical result indicates that this convergence is quadratic.

Having shown the validation of the construction of $\Xi$, we determine a proper upper bound for the Casimir integration \eqref{slp and matrix} by Remark \ref{Determine the upperbound}. 
With the upper bound of the integration determined, one can start to estimate the Casimir energy between two spheres with radius $r_1 = r_2 = 1$ at the 
distance of $Z$ via the formula \eqref{KSSF and CasE} in two different refinement levels: $h_{\text{fine}} = 0.05$ 
($\text{dim}(\mathsf{V}_{\mathrm{i}k}) = 12603$) and $h_{\text{coarse}} = 0.1$ ($\text{dim}(\mathsf{V}_{\mathrm{i}k}) = 3192$). Afterwards, the extrapolation result can be computed by these Casimir energy estimates.

\begin{figure}[H]
    \centering
    \includegraphics[width = \textwidth]{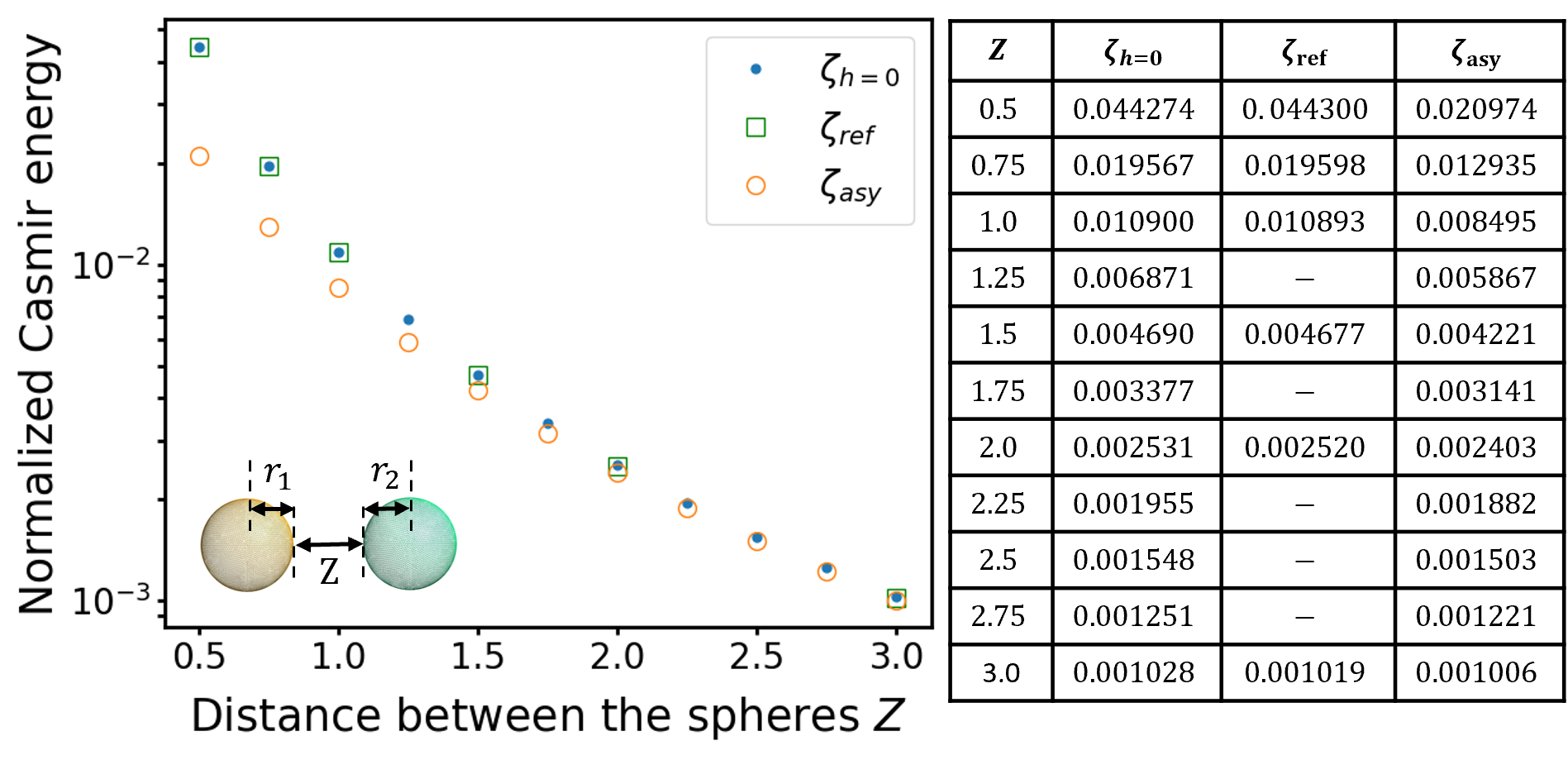}
    \caption[Caption for LOF]{{\color{gray}Left:} Negative normalized Casimir energy  \protect\footnotemark computed by the extrapolation (blue circle), asymptotic series (orange hollow circle) in two spheres with equal radii's case. The radius is $r_{1} = r_{2} = 1$ and the distance $Z$ 
    ranges from 0.5 to 3.0. The green hollow square represents the data of \cite{kenneth2008casimir}. {\color{gray}Right:} The table lists all the relevant data values in the figure.}
    \label{Casimir energy between spheres with equal radii}
\end{figure}

According to \cite{emig2008casimir}, the Casimir energy between two spheres (with equal radii $r$) at asymptotically 
large separations can be obtained as a series in terms of the ratio of centre distance $L$ ($l = 2r + Z$) to sphere radius $R$:
\begin{align}\label{Asymptotic equal radii}
   \zeta_{\text{asy}} = -\frac{\hbar c}{\pi}\frac{1}{L}\sum_{n=0}^{\infty}b_{n}\left(\frac{r}{l}\right)^{n+2},
\end{align}
where the first six coefficients are 
$b_{0} = -1/4$, $b_{1} = -1/4$,  $b_{2} = -77/48$,  $b_{3} = -25/16$,  $b_{4} = -29837/2880$, $b_{5} = -6491/1152$. Figure 
\ref{Casimir energy between spheres with equal radii} shows the comparison between the Casimir energy computed from asymptotic series 
\eqref{Asymptotic equal radii} and the exact value evaluated through Richardson extrapolation and the reference value $\zeta_{\text{ref}}$ provided in \cite[Equation (64)]{kenneth2008casimir}. Here, we observe that the asymptotic value gradually 
approaches to the exact value as the distance $Z$ increases since the asymptotic expansion \eqref{Asymptotic equal radii} only works when the distance 
between two spheres is asymptotically large.

\footnotetext{The negative normalized Casimir energy is $-\xi/\hbar c$, for $\xi$ defined in \eqref{KSSF and CasE}. Note that for the labels in all the figures,
the normalized Casimir energy means the negative normalized.}

\begin{figure}[H]
    \centering
    \includegraphics[scale = 0.5]{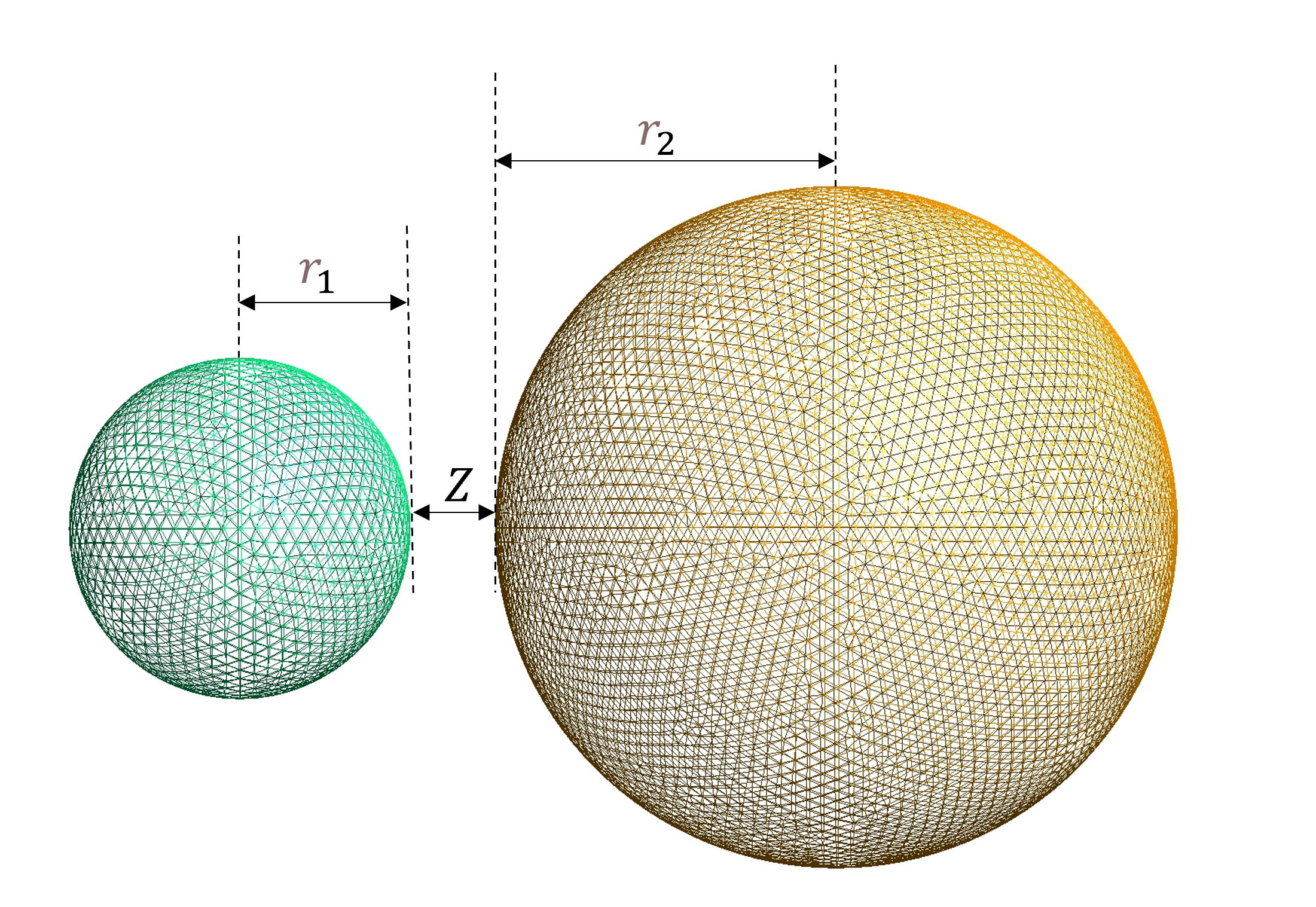}
    \caption{Two spheres with unequal radii $r_{1} = 0.5$ and $r_{2} = 1$ and $Z$ is the minimal distance between them.\\
    \hspace*{1.5cm}$h_{\text{coarse}} = 0.1$: $\text{dim}(\mathrm{V}_{\mathrm{i}k}) = 2023$,  N\textsuperscript{\underline{o}} of elements on both grids $ = 4038$;\\
    \hspace*{1.5cm}$h_{\text{fine}} = 0.05$: $\text{dim}(\mathrm{V}_{\mathrm{i}k}) = 7891$,  N\textsuperscript{\underline{o}} of elements on both grids $ = 15774$}
    \label{Two spheres with unequal radii}
\end{figure}

Now, let us consider the case when two spheres have different radii $r_{1}$, $r_{2}$ (see Figure \ref{Two spheres with unequal radii}). 

In this case, one can still determine the upper bound of the integration by fitting the integrand function curve and considering the error tolerance.
Afterwards, we would like to compare the extrapolation value of the Casimir energy computed through the Richardson extrapolation with the asymptotic expansion. By denoting the centre distance as
$l = r_{1} + r_{2} + Z$, the asymptotic series of the Casimir energy between these two spheres is written by
\begin{align}\label{Asymptotic unequal radii}
    \zeta_{\text{asy}} = -\frac{\hbar c}{\pi}\frac{1}{l}\sum_{n=0}^{\infty}\tilde{b}_{n}(\eta)\left(\frac{r_{1}}{L}\right)^{n+2},
\end{align}
where the coefficients $\{\tilde{b}_{n}\}$ depend on the parameter $\eta = r_{2}/r_{1}$ and the first six coefficients are
\begin{align*}
    \tilde{b}_{0} &= -\frac{\eta}{4}, \ \ \ \ \ \tilde{b}_{1} = -\frac{\eta + \eta^{2}}{8}, \ \ \ \ \  \tilde{b}_{2} = -\frac{34(\eta+\eta^{3})+ 9\eta^{2}}{48}, \ \ \ \ \ \tilde{b}_{3} = -\frac{2(\eta+\eta^{4}) + 23(\eta^{2} + \eta^{3})}{32}, \\ 
    \tilde{b}_{4} &= -\frac{8352(\eta + \eta^{5})+ 1995(\eta^{2} + \eta^{4}) + 38980\eta^{3}}{5760}, \ \ \ \ \ \tilde{b}_{5} = -\frac{-1344(\eta+\eta^{6}) + 5478(\eta^{2} + \eta^{5})+2357(\eta^{3} + \eta^{4})}{2304}.
\end{align*}

In the following experiment, the radii of the spheres shown in Figure \ref{Two spheres with unequal radii} are set as $r_{1} = 0.5$ and $ r_{2} = 1$. 
As in the previous example, the exact value of the Casimir energy is computed through the Richardson extrapolation and the coarse and fine grid size are $h_{\text{fine}} = 0.05$ ($\text{dim}(\mathsf{V}_{\mathrm{i}k}) = 7893)$ and 
$h_{\text{coarse}} = 0.1$ ($\text{dim}(\mathsf{V}_{\mathrm{i}k}) = 2023)$, separately. 

\begin{figure}[H]
    \centering
    \includegraphics[width = \textwidth]{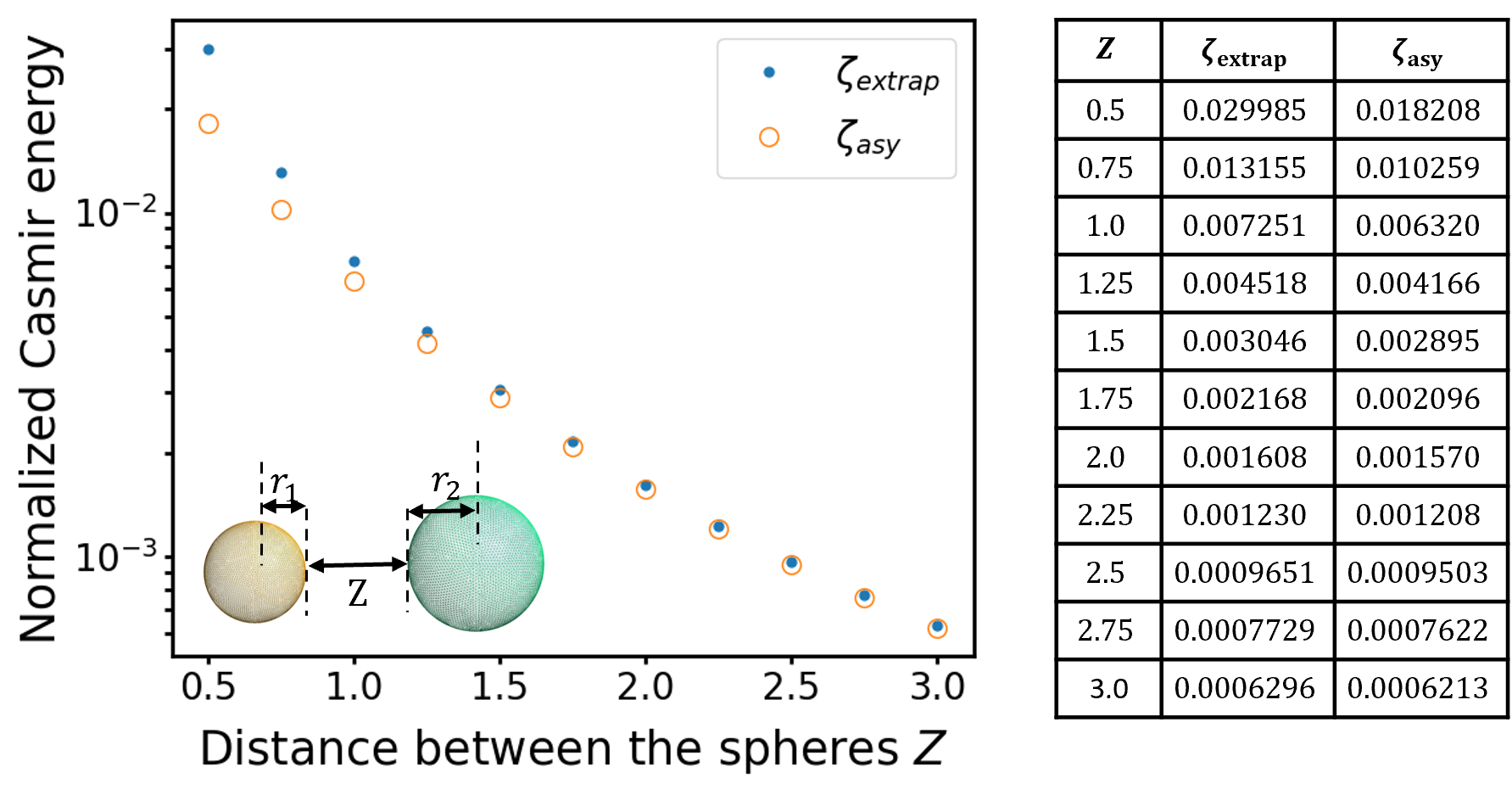}
    \caption{Negative normalized Casimir energy in two spheres with unequal radii's case. The radius is $R = 1$ and the distance $Z$ 
    ranges from 0.5 to 3.0. The exact value of the (negative normalized) Casimir energy has been written 
    beside the data point, which is round up to 4 significant digits.}
    \label{Casimir energy between spheres with unequal radii}
\end{figure}

In this case, the asymptotic value of the Casimir 
energy was estimated by the series \eqref{Asymptotic unequal radii} and the comparison between the extrapolation value and asymptotic one is shown in Figure 
\ref{Casimir energy between spheres with unequal radii}. Again, one can notice that when the distance between two spheres decreases, the asymptotic value gets 
close to the extrapolation one.

\begin{figure}[H]
\centering
\begin{minipage}{.65\textwidth}
  \centering
  \includegraphics[width=\textwidth]{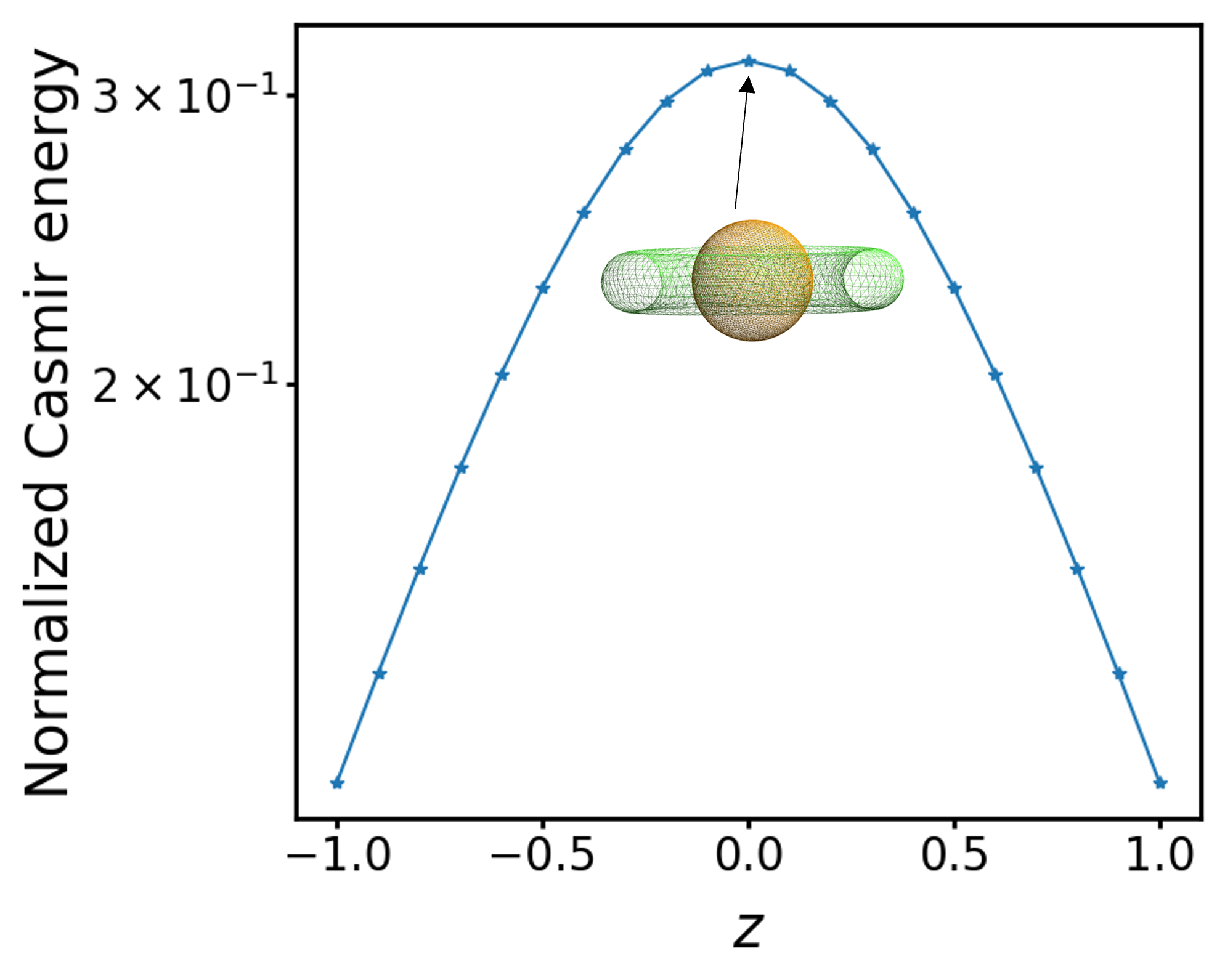}
\end{minipage}%
\begin{minipage}{.35\textwidth}
  \centering
  \vspace*{-0.6cm}
  \includegraphics[width=\textwidth]{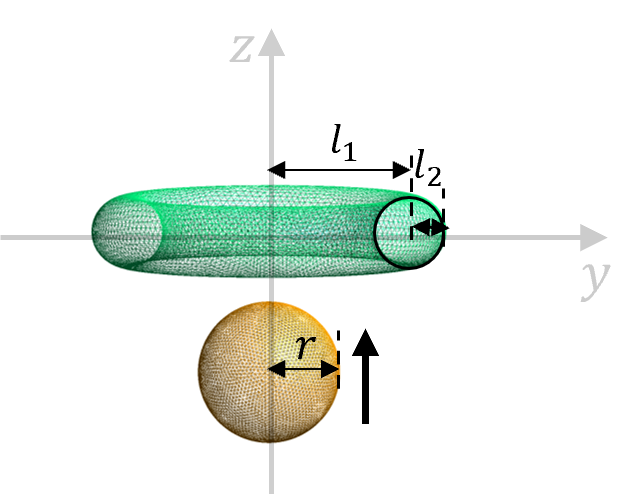}
%   \caption{A subfigure}
%   \label{fig:Sphere_torus}
\end{minipage}
\caption{Negative normalized Casimir energy between a torus and a sphere when the sphere moves along the $z$-axis. The parameters of the torus are $l_1 = 2$, $l_2 = 0.5$ and the radius of the sphere is $r = 1$.}
\label{fig:Torus_sphere_CasE}
\end{figure}

After showing the validation of the numerical framework for computing the Casmir energy, we would like to end this section with computing the negative normalized Casimir energy between one torus and one sphere. For the torus, it is centering at the origin and the distance from the center of the tube to the center of the torus is $l_1 = 2$ and the radius of the tube is $l_2 = 0.5$; for the sphere, 
it has radius $r = 1$ and its center is always on the $z$-axis (see Figure \ref{fig:Torus_sphere_CasE} (Right)). By  Figure \ref{fig:Torus_sphere_CasE} (Left), one can see that when the sphere and the torus share the same center, the negative normalized Casimir energy has the largest magnitude.

\subsection{Realistic objects case}
In this part, the Casimir energy between the objects with special shapes such as the menger sponges, ice crystals and ellipsoids will be computed 
through the Richardson extrapolation mentioned in the beginning of this section and the values labelled in the following figures are accurate within three significant digits. 
Note that the matrix size of the involved matrix in each example has been stated in the figures.

\begin{figure}[H]
    \centering
    \captionsetup[subfigure]{justification=centering}
    \subfloat[Level 0 \\ $h_{\text{coarse}} = 0.1$: $\text{dim}(\mathsf{V}_{\mathrm{i}k}) = 1456$,  N\textsuperscript{\underline{o}} of elements on both grids $ = 2904$
    \\ $h_{\text{fine}} = 0.05$: $\text{dim}(\mathsf{V}_{\mathrm{i}k}) = 5664$,  N\textsuperscript{\underline{o}} of elements on both grids $ = 11120$]{{\includegraphics[scale = 0.5]{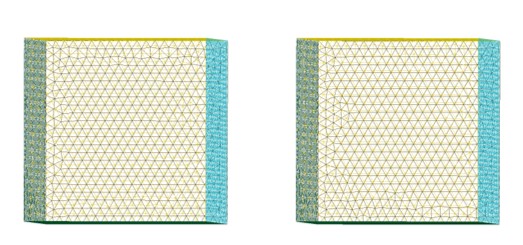} }}
    \qquad
    \subfloat[Level 1 \\ $h_{\text{coarse}} = 0.1$: $\text{dim}(\mathsf{V}_{\mathrm{i}k}) = 3092$,  N\textsuperscript{\underline{o}} of elements on both grids $ = 6216$
    \\ $h_{\text{fine}} = 0.05$: $\text{dim}(\mathsf{V}_{\mathrm{i}k}) = 8510$,  N\textsuperscript{\underline{o}} of elements on both grids $ = 17052$]{{\includegraphics[width=0.4\textwidth]{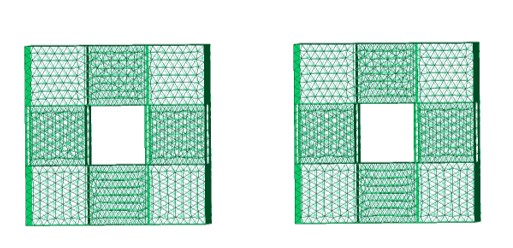} }}
    \qquad
    \subfloat[Level 2 \\ $h_{\text{coarse}} = 0.1$: $\text{dim}(\mathsf{V}_{\mathrm{i}k}) = 14464$,  N\textsuperscript{\underline{o}} of elements on both grids $ = 29568$
    \\ $h_{\text{fine}} = 0.05$: $\text{dim}(\mathsf{V}_{\mathrm{i}k}) = 27136$,  N\textsuperscript{\underline{o}} of elements on both grids $ = 54912$]{{\includegraphics[width=0.5\textwidth]{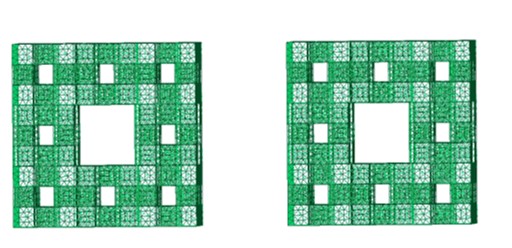} }}
    \caption{Menger sponges in different levels. The length of each sponge is 1.}
    \label{Menger sponges}
\end{figure}

Figure \ref{Menger sponges} plots the menger sponges in different levels $(0, 1 $ and $ 2)$ and the length of these sponges is always 1. Afterwards, the Casimir 
energy between two menger sponges in the same level are listed in Table \ref{Negative normalized Casimir energy in two menger sponges' case}. 
In addition, inside the extrapolation process, when $h_{\text{fine}} = 0.05$, the $\text{dim}(\mathsf{V}_{\mathrm{i}k}) = 5664$, 8510 and 27136 and 
when $h_{\text{coarse}} = 0.1$, the $\text{dim}(\mathsf{V}_{\mathrm{i}k}) = 1456$, 3092 and 14464 in different level (0, 1 and 2) cases, separately. 
By comparing the data 
point in this table, it is easy to find that the Casimir energy decreases as the number of the iteration increases since the cross-sectional 
area gets smaller.

\begin{table}[H]
    \centering
    \begin{tabular}{ |P{2cm}||p{2cm}|p{2cm}|p{2cm}|  }
        \hline
        \multicolumn{4}{|c|}{Negative normalized Casimir energy in two menger sponges' case} \\
        \hline
        Distance & Level 0 & Level 1 & Level 2\\
        \hline
        0.5   & 0.08350    & 0.08229     & 0.08112\\
        0.75  & 0.02737    & 0.02688     & 0.02670\\
        1.0   & 0.01305    & 0.01288     & 0.01282\\
        1.25  & 0.007357   & 0.007283    & 0.007252\\
        1.5   & 0.004607   & 0.004568    & 0.004551\\
        1.75  & 0.003099   & 0.003076    & 0.003065\\
        2.0   & 0.002195   & 0.002181    & 0.002174\\
        2.25  & 0.001618   & 0.001608    & 0.001603\\
        2.5   & 0.001230   & 0.001223    & 0.001220\\
        2.75  & 0.0009593  & 0.0009541   & 0.0009514\\
        3.0   & 0.0007638  & 0.0007598   & 0.0007577\\
        \hline
       \end{tabular}
       \caption{\label{Negative normalized Casimir energy in two menger sponges' case} Negative normalized Casimir energy in two menger sponges' case}
    \end{table}

\begin{figure}[H]
    \centering
    \includegraphics[scale = 1]{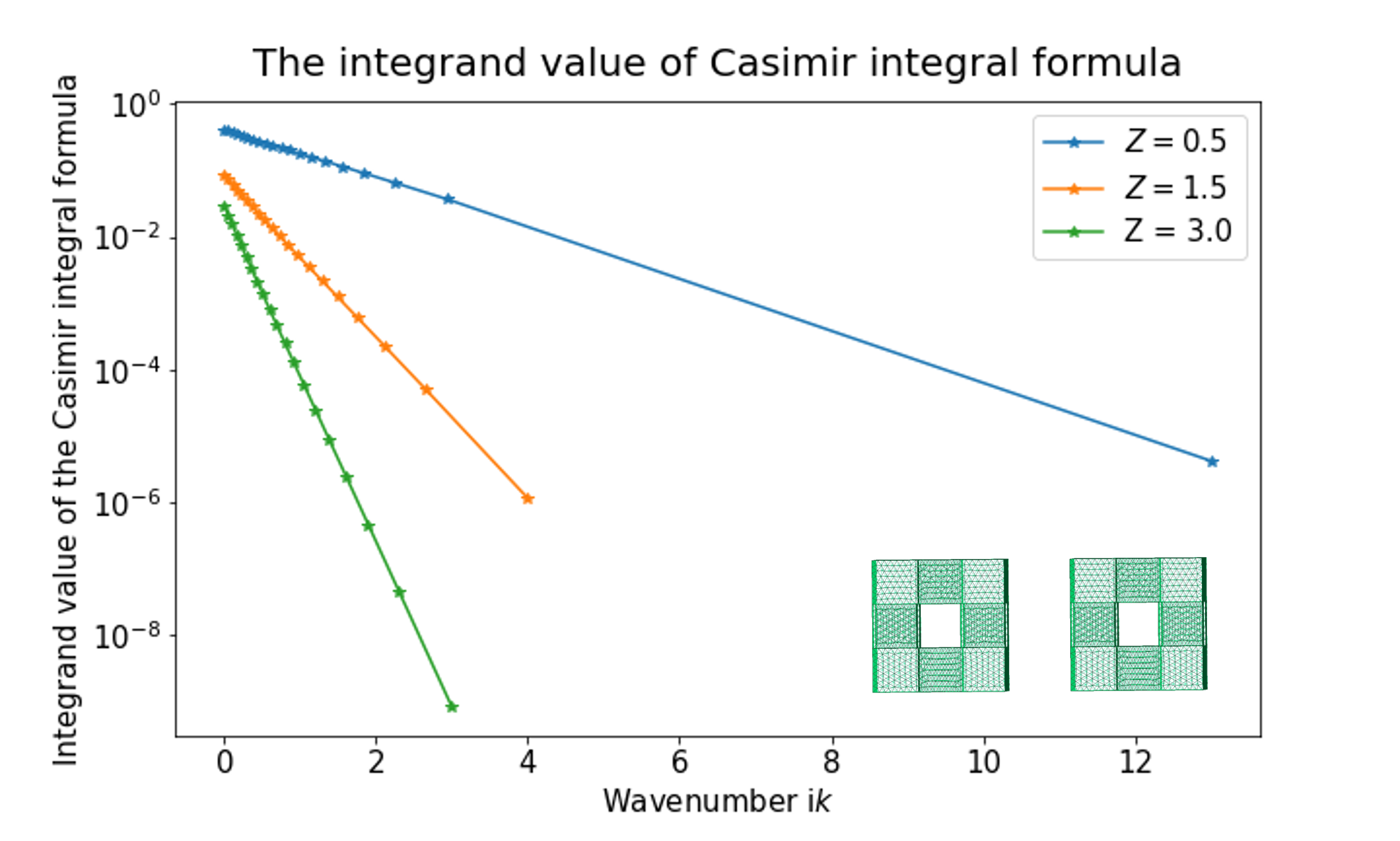}
    \caption{The integrand of the Casimir energy between two menger sponges in Level 1 with distance $Z = 0.5$, 1.5 and 3.0.}
    \end{figure}

\begin{figure}[H]
        \centering
        \includegraphics[width = \textwidth]{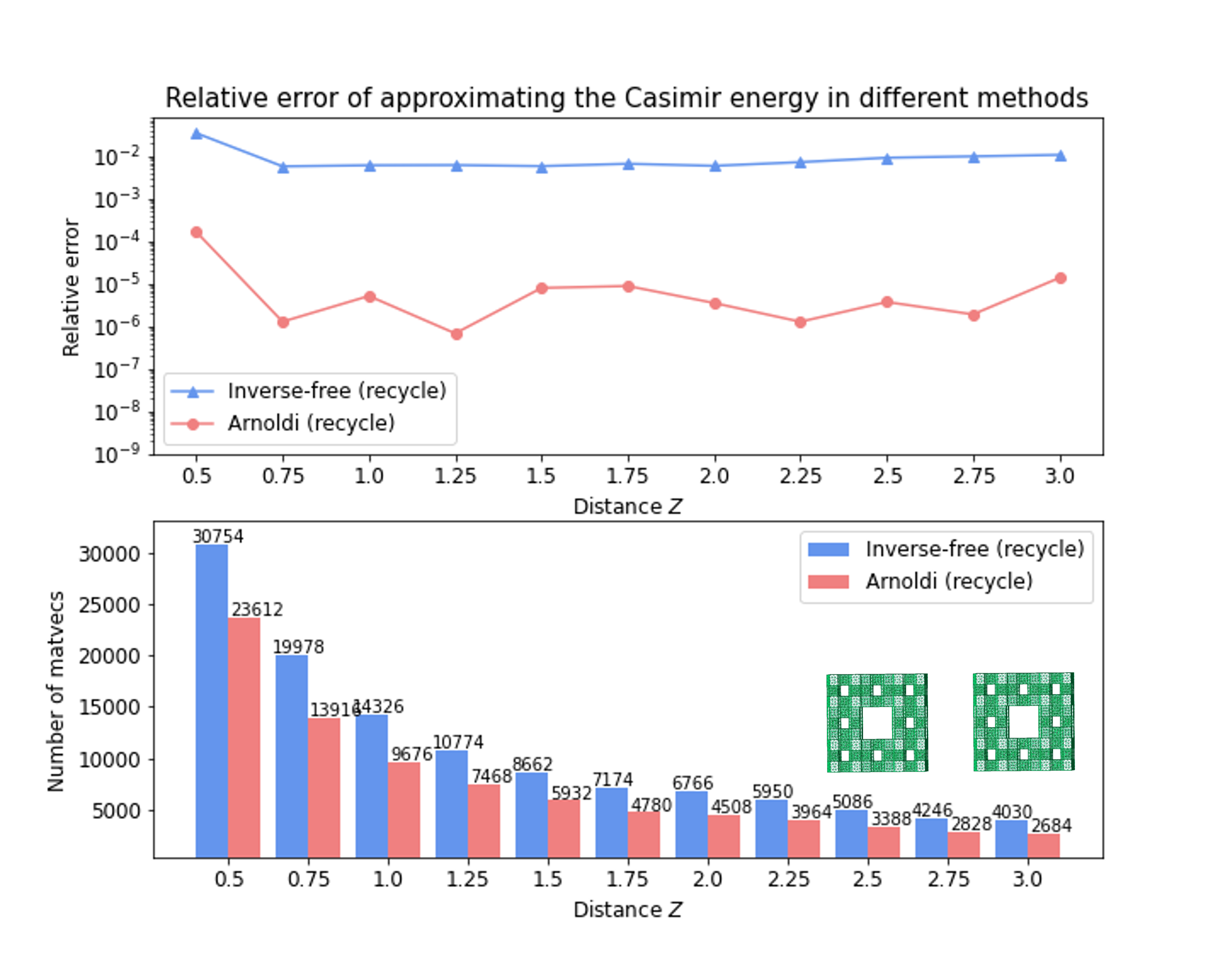}
        \caption{Menger sponges in Level 2's case: relative distance between the reference value (computed by Richardson extrapolation) with the estimates evaluated from the standard Arnoldi 
        method with subspace recycled (solid red circles) and inverse-free Krylov subspace method 
        with subspace recycled (solid blue triangles). The dimension of the Krylov subspace is $m = 100$. The recycled eigenvectors have the corresponding eigenvalue 
        whose logarithm is larger than $10^{-5}$.}
\end{figure}

\begin{figure}[H]
    \begin{subfigure}{0.3\linewidth}
        \centering
        \includegraphics[scale = 0.4]{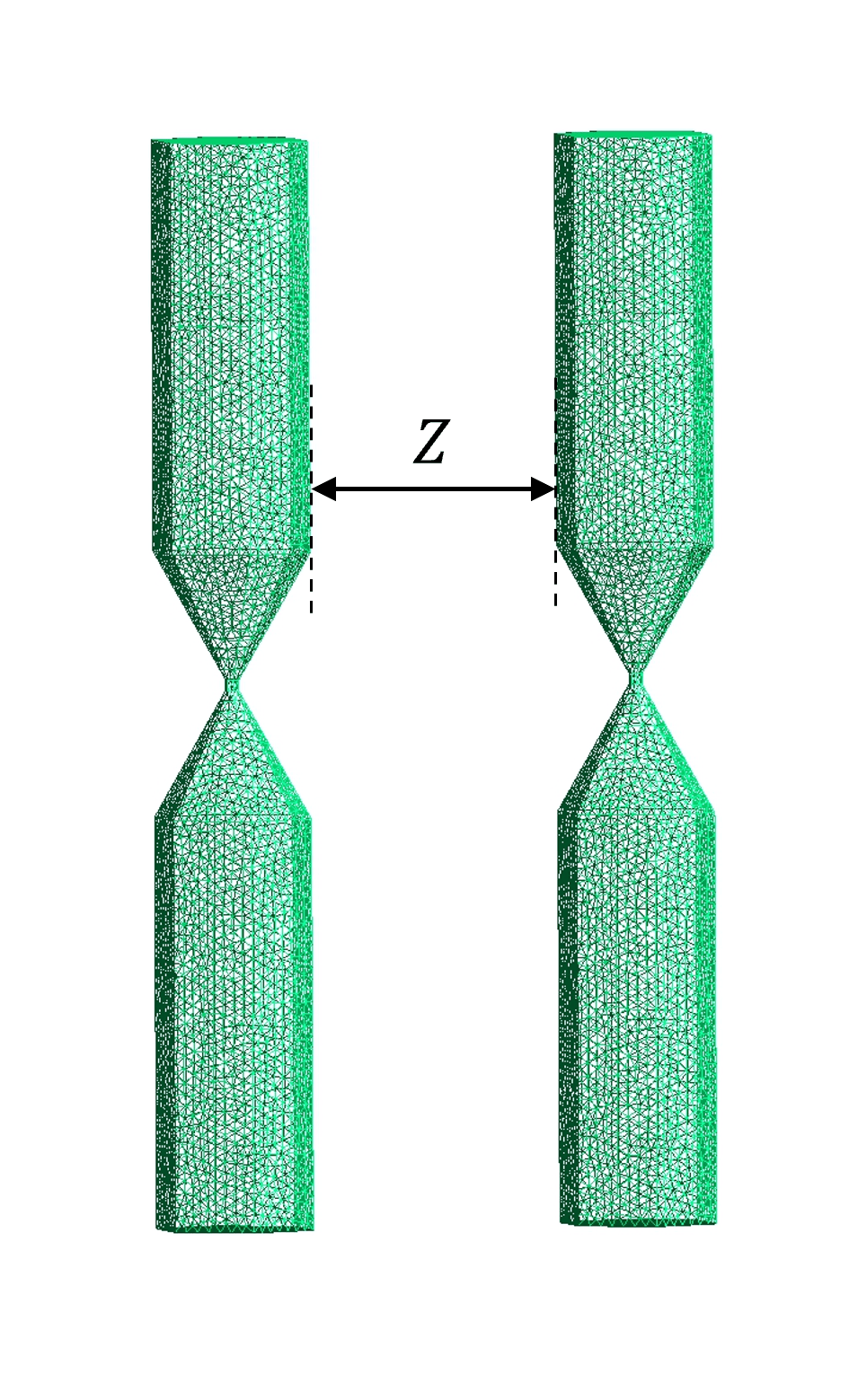}
        \caption{Two branches: $\text{dim}(\mathsf{V}_{\mathrm{i}k}) = 8792$ \newline N\textsuperscript{\underline{o}} of elements on both grids $ = 17576$}
        \end{subfigure}
        \begin{subfigure}{0.3\linewidth}
            \centering
            \includegraphics[scale = 0.4]{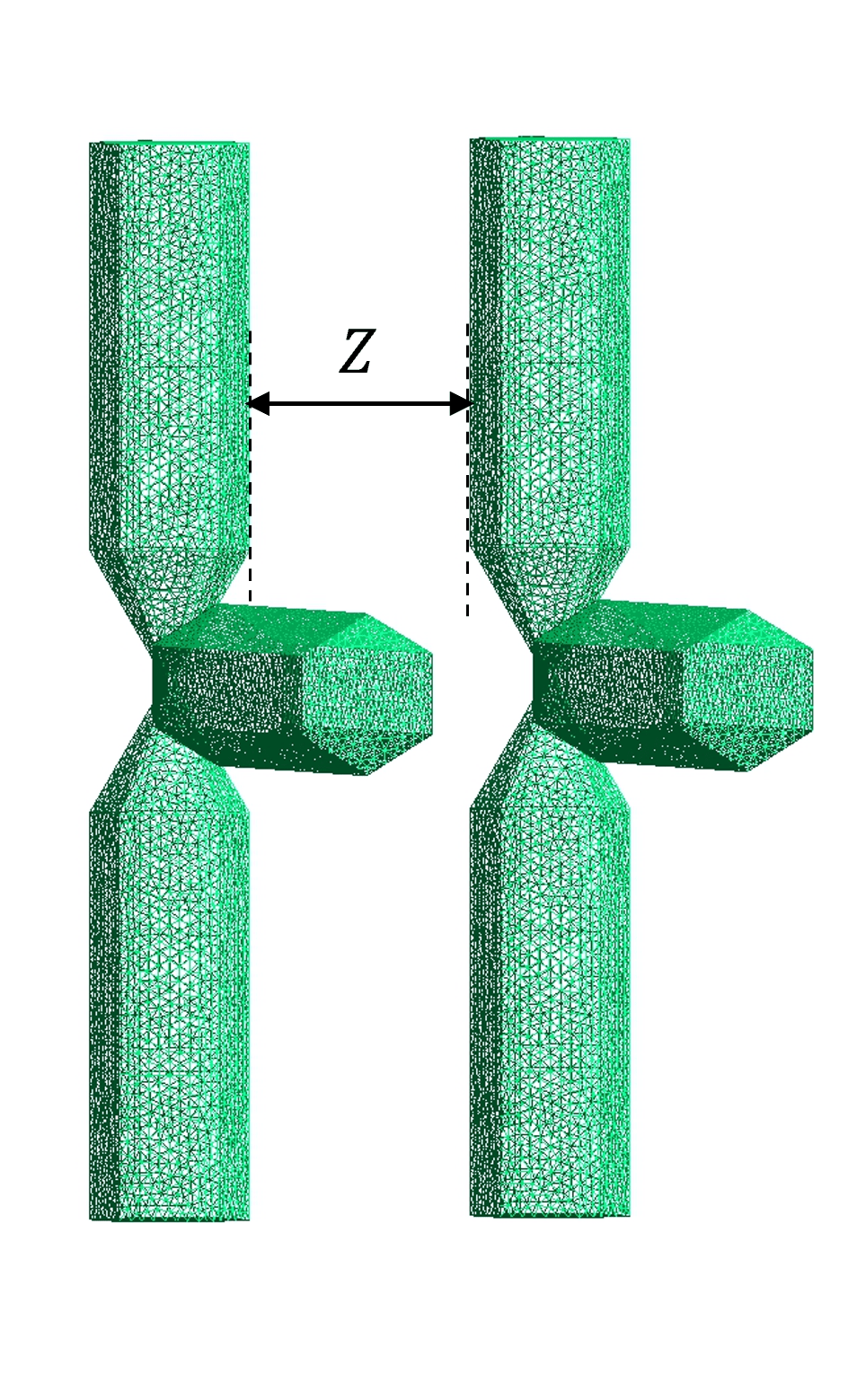}
            \caption{Three branches: $\text{dim}(\mathsf{V}_{\mathrm{i}k}) = 13104$ \newline N\textsuperscript{\underline{o}} of elements on both grids $ = 26200$}
            \end{subfigure}
            \begin{subfigure}{0.3\linewidth}
                \centering
                \includegraphics[scale = 0.4]{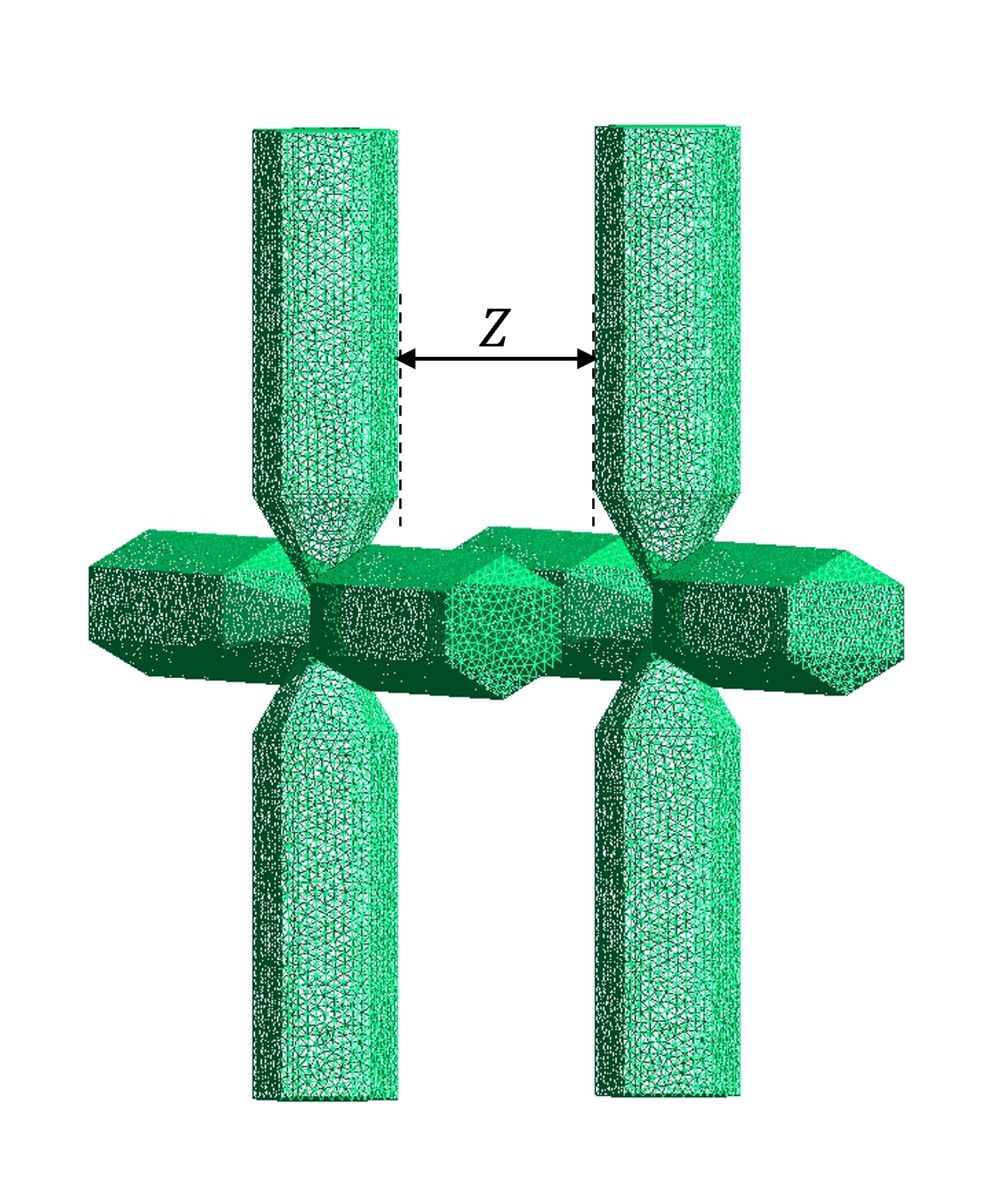}
                \caption{Four branches: $\text{dim}(\mathsf{V}_{\mathrm{i}k}) = 17554$ \newline N\textsuperscript{\underline{o}} of elements on both grids $ = 35100$}
                \end{subfigure}\\[1ex]
    %\begin{subfigure}{.5\linewidth}
    \centering
    \captionsetup[subfigure]{oneside,margin={0.4cm,0cm}}
    \subfloat[Five branches: $\text{dim}(\mathsf{V}_{\mathrm{i}k}) = 21950$ \newline N\textsuperscript{\underline{o}} of elements on both grids $ = 43900$]{\includegraphics[scale=0.4]{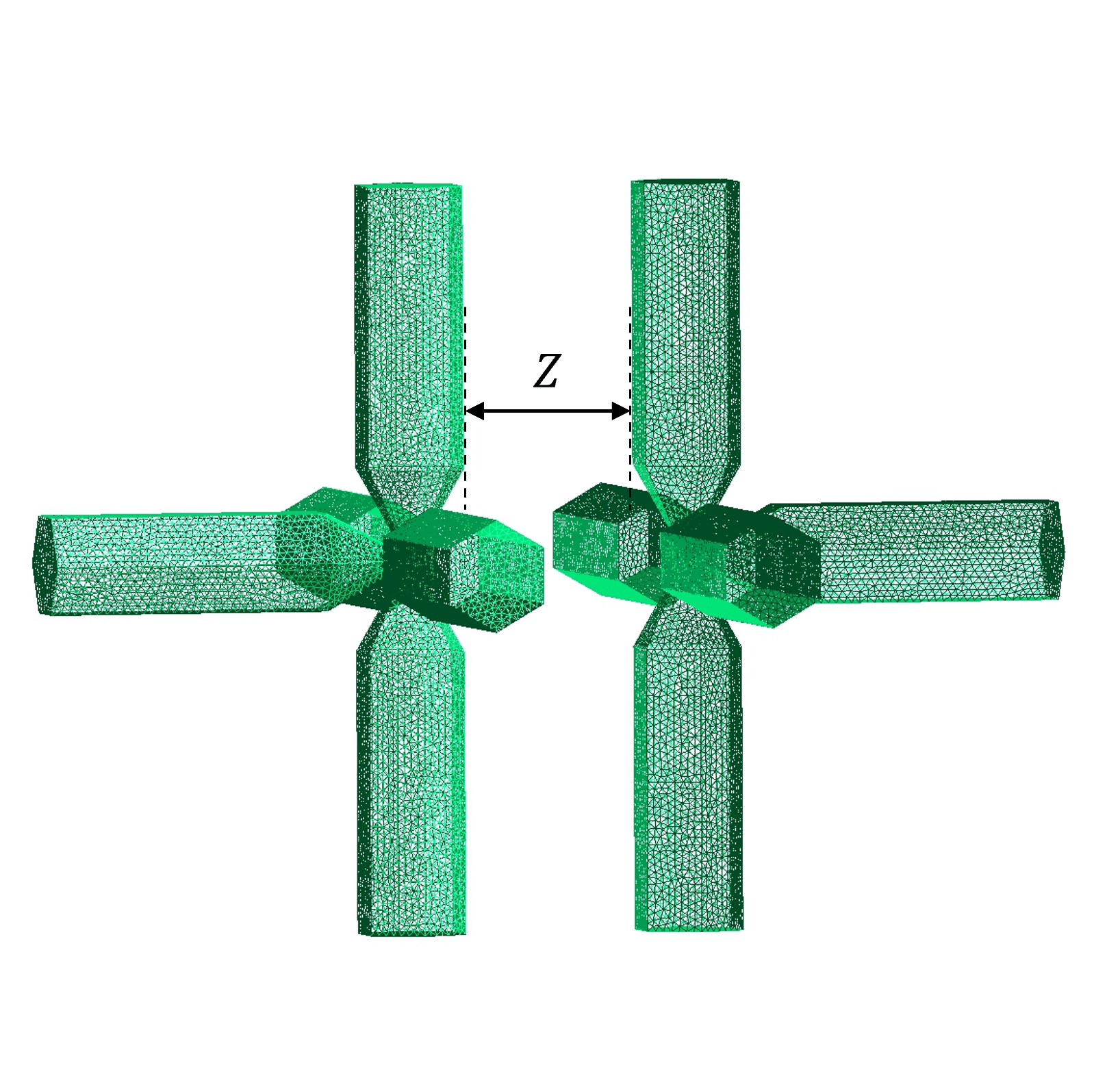}}
    \centering
    \hspace*{1.5cm}
    \captionsetup[subfigure]{oneside,margin={1.2cm,0cm}}
    \subfloat[Six branches: $\text{dim}(\mathsf{V}_{\mathrm{i}k}) = 26262$ \newline N\textsuperscript{\underline{o}} of elements on both grids $ = 52556$]{\includegraphics[scale=0.4]{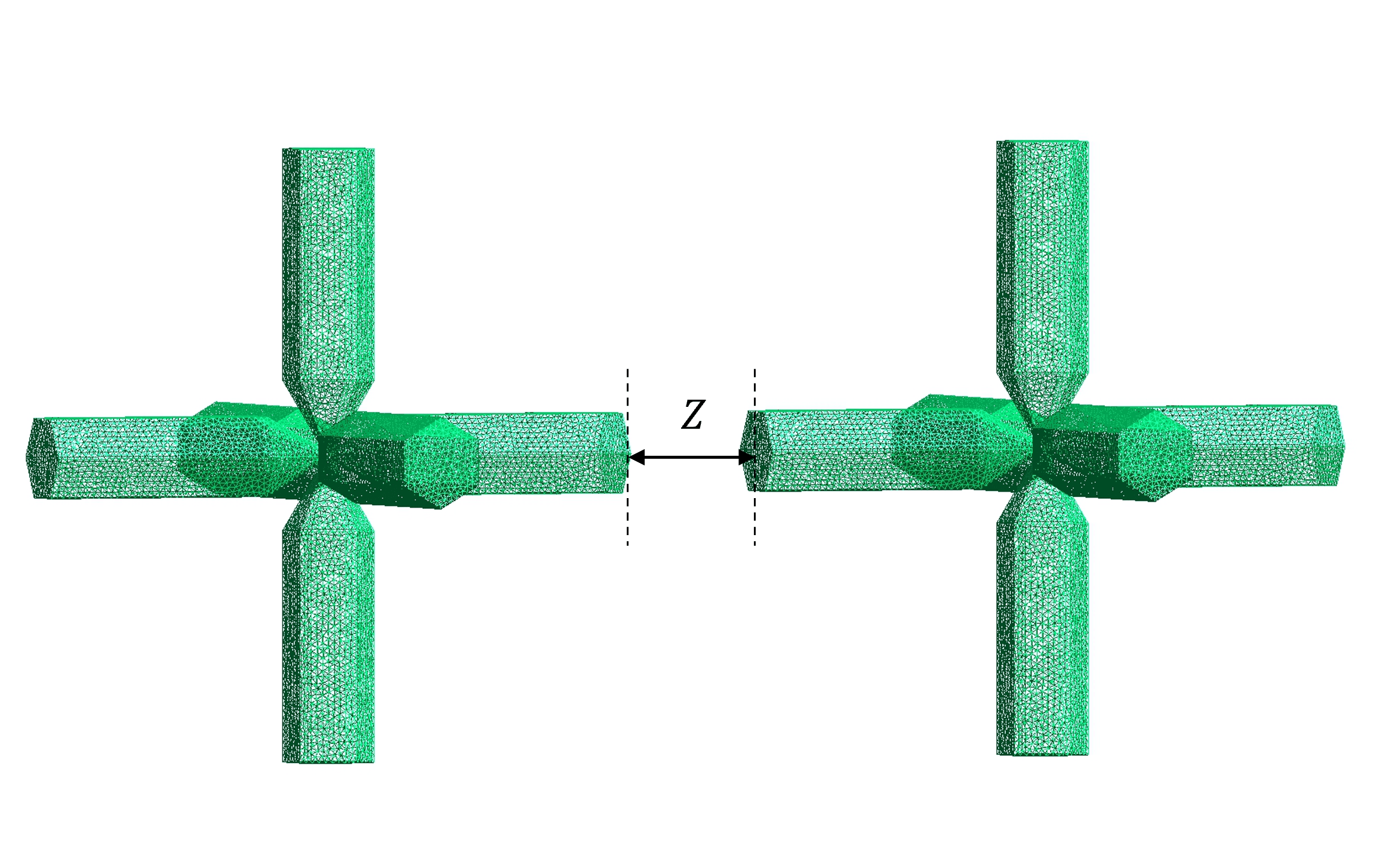}}
    \caption{Ice crystals with different number of branches}
    \label{Ice crystals with different number of branches}
    \end{figure}

%==========================================================================================
In the next example, the scatterers are ice crystals with different number of branches ranging from 2 to 6 (see Figure \ref{Ice crystals with different number of branches}).

    \begin{table}[H]
        \centering
        \begin{tabular}{ |P{2cm}||p{2cm}|p{2cm}|p{2cm}|p{2cm}|p{2cm}|  }
            \hline
            \multicolumn{6}{|c|}{Negative normalized Casimir energy in ice crystals' case} \\
            \hline
            Distance & 2-branches & 3-branches & 4-branches & 5-branches & 6-branches\\
            \hline
            0.5   & 0.04112    & 0.05989    & 0.07848   & 0.07873    & 0.01128\\
            0.75  & 0.01499    & 0.02184    & 0.02855   & 0.02873    & 0.005017\\
            1.0   & 0.007403   & 0.01080    & 0.01412   & 0.01428    & 0.002965\\
            1.25  & 0.004242   & 0.006198   & 0.008113  & 0.008242   & 0.001985\\
            1.5   & 0.002672   & 0.003905   & 0.005117  & 0.005223   & 0.001427\\
            1.75  & 0.001797   & 0.002624   & 0.003442  & 0.003530   & 0.001074\\
            2.0   & 0.001268   & 0.001849   & 0.002428  & 0.002501   & 0.0008357\\
            2.25  & 0.0009288  & 0.001353   & 0.001776  & 0.001839   & 0.0006664\\
            2.5   & 0.0007007  & 0.001019   & 0.001338  & 0.001391   & 0.0005410\\
            2.75  & 0.0005413  & 0.0007863  & 0.001033  & 0.001078   & 0.0004469\\
            3.0   & 0.0004270  & 0.0006188  & 0.0008134 & 0.0008526  & 0.0003741\\
            \hline
           \end{tabular}
           \caption{\label{Negative normalized Casimir energy in ice crystals' case table} Negative normalized Casimir energy in 2- to 6-branched ice crystals' case}
        \end{table}

        \begin{figure}[H]
            \centering
            \includegraphics[scale = 1]{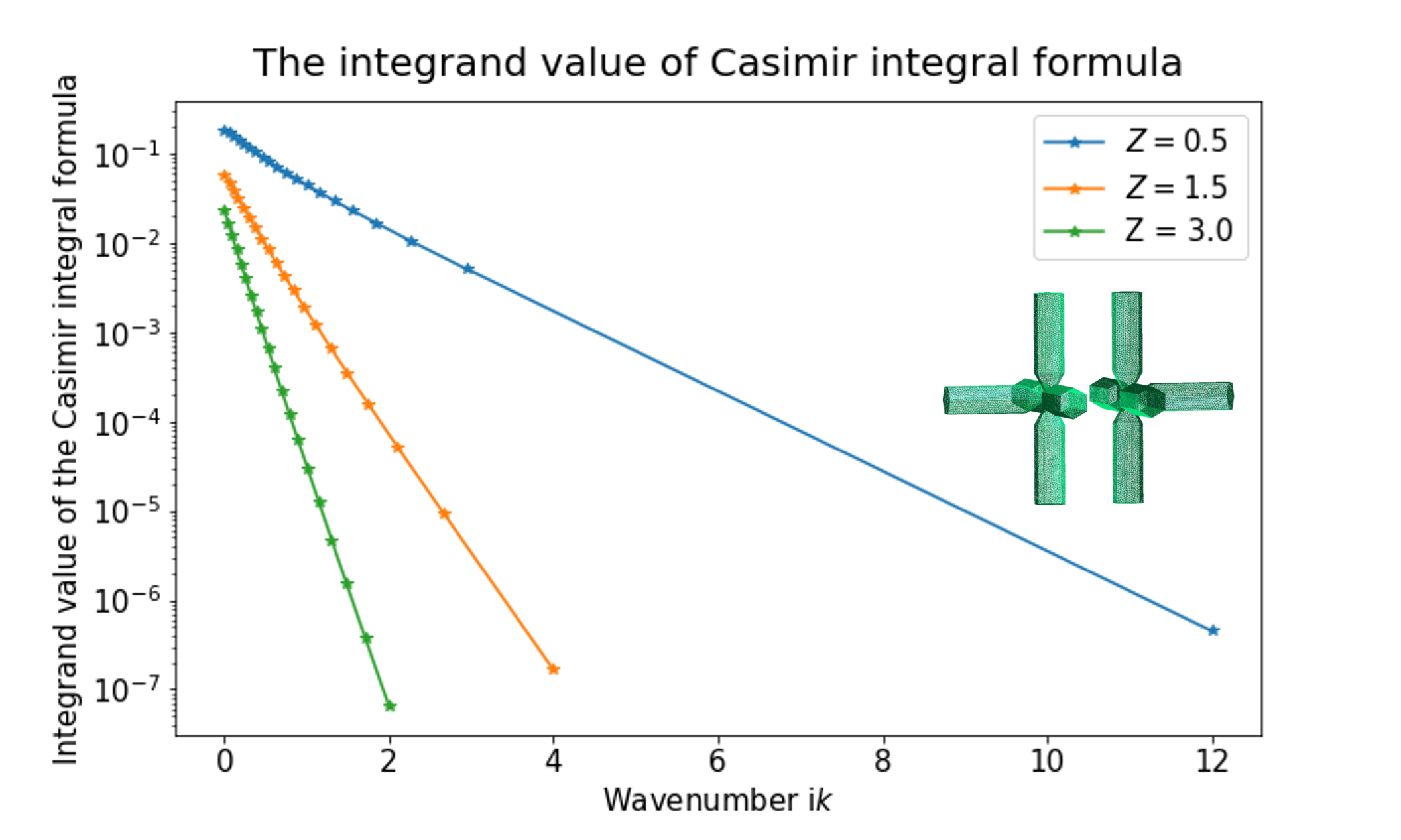}
            \caption{The integrand of the Casimir energy between two five-branches ice crystals with distance $Z = 0.5$, 1.5 and 3.0.}      
              \end{figure}
        
        \begin{figure}[H]
                \centering
                \includegraphics[width = \textwidth]{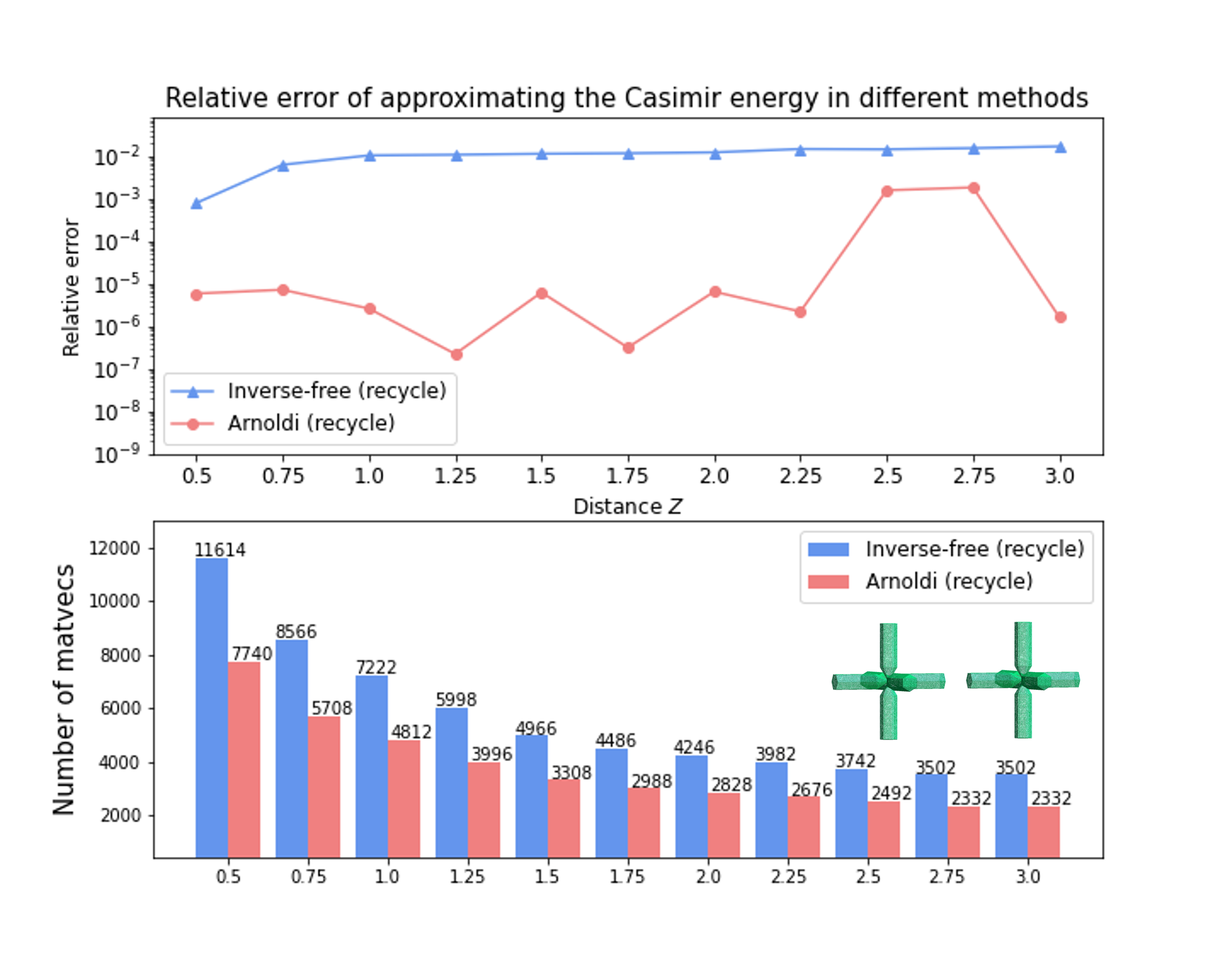}
                \caption{Six-branches ice crystals' case: relative distance between the reference value (computed by Richardson extrapolation) with the estimates evaluated from the standard Arnoldi 
                method with subspace recycled (solid red circles) and inverse-free Krylov subspace method with subspace recycled (solid blue triangles). The dimension of the Krylov subspace is $m = 100$.
                The recycled eigenvector has the corresponding eigenvalue whose logarithm is larger than $10^{-5}$.}       
             \end{figure}

It is not hard to imagine that the Casimir energy would be different when rotating the scatterers and keeping the distance between them unchanged. Therefore, 
in the last example, we would see how the Casimir energy between two identical ellipsoids changes as one of the ellipsoids rotates.

\begin{figure}[H]
    \begin{subfigure}{\linewidth}
        \centering
        \includegraphics[scale = 0.3]{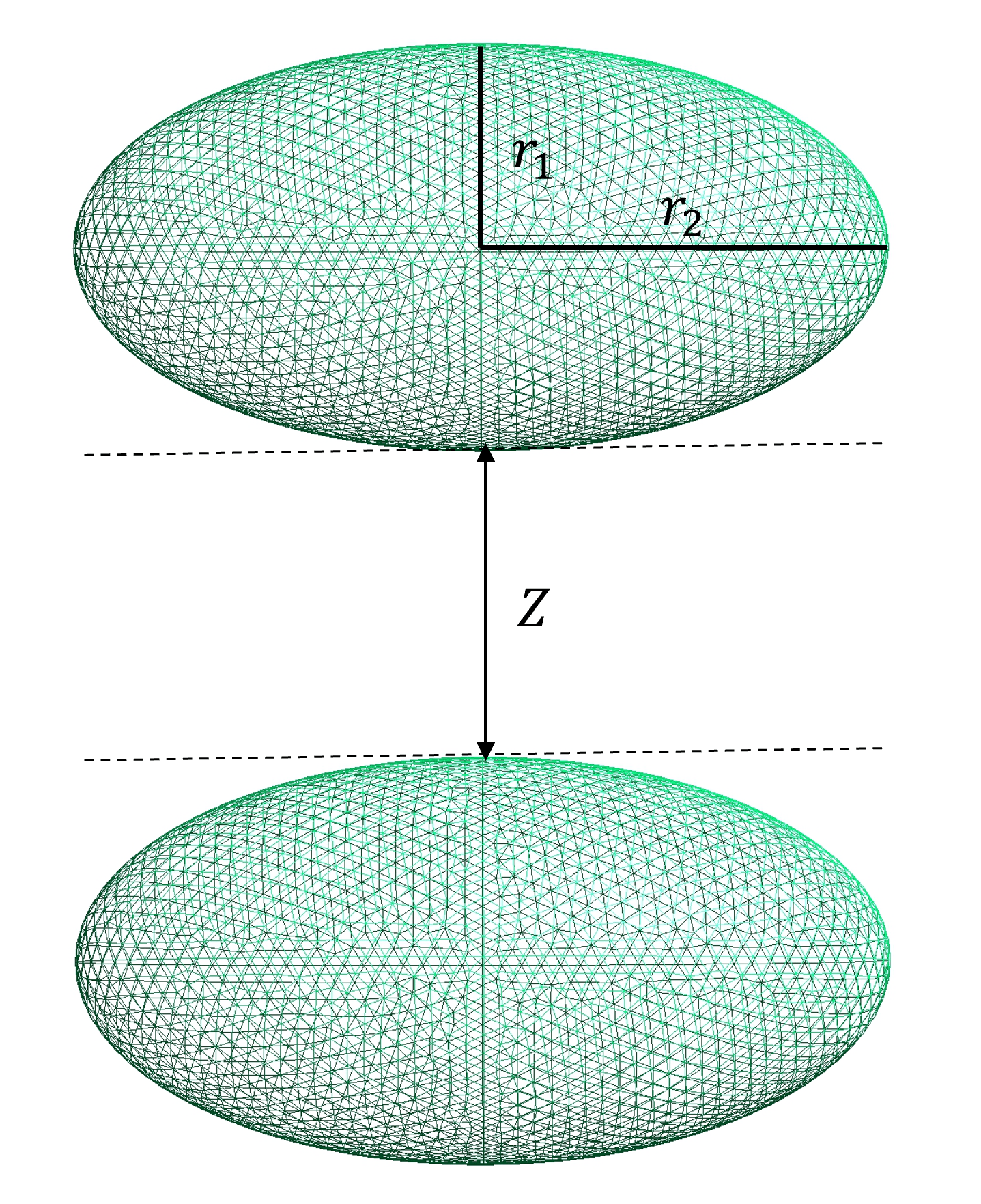}
        \caption{Without rotation}
        \label{Without rotation}
        \end{subfigure}\\[1ex]
    \begin{subfigure}[t]{.5\linewidth}
    \centering
    \includegraphics[scale = 0.3]{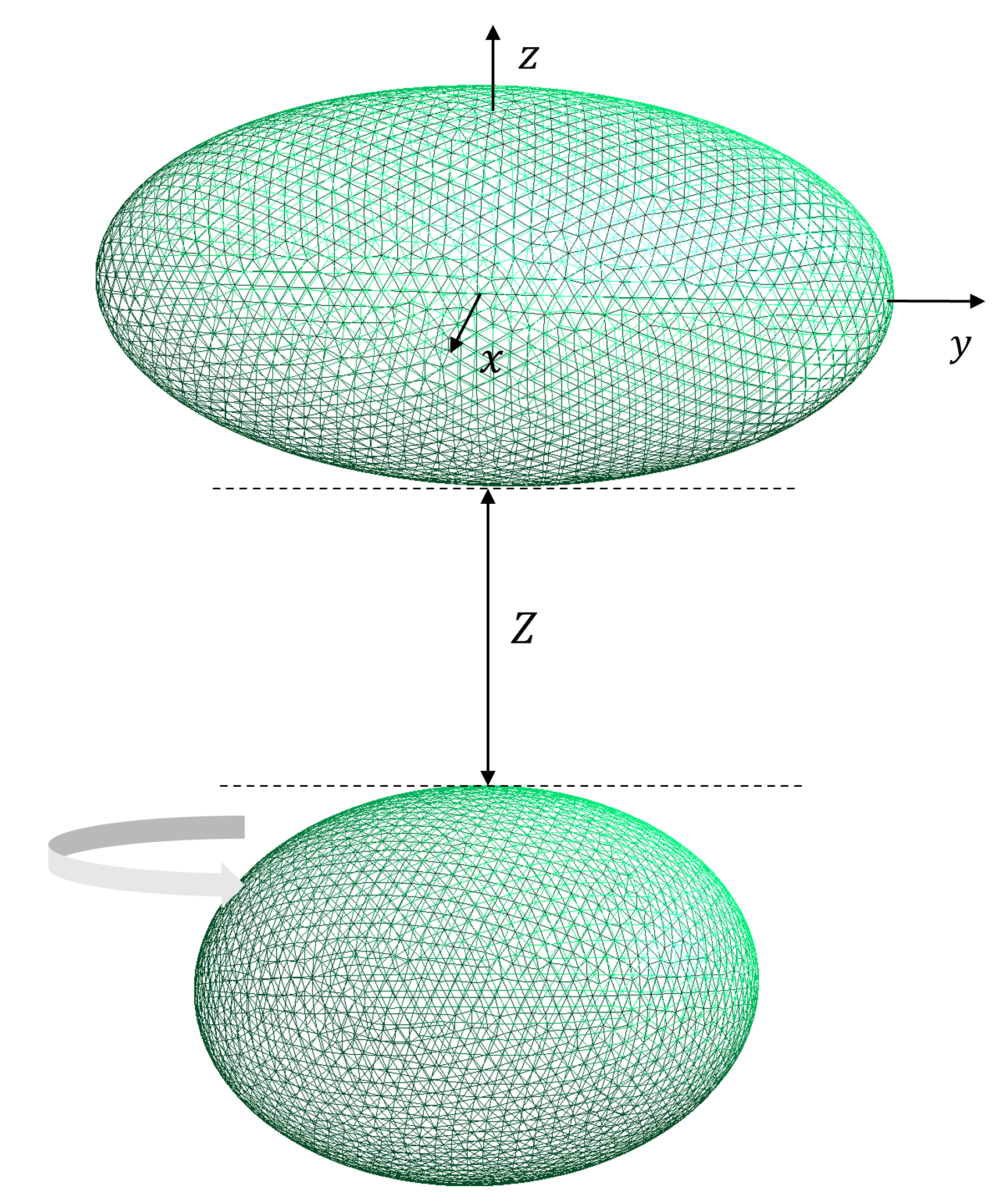}
    \caption{Rotation around z-axis}
    \label{Rotation around z-axis}
    \end{subfigure}%
    \begin{subfigure}[t]{.5\linewidth}
    \centering
    \includegraphics[scale = 0.07]{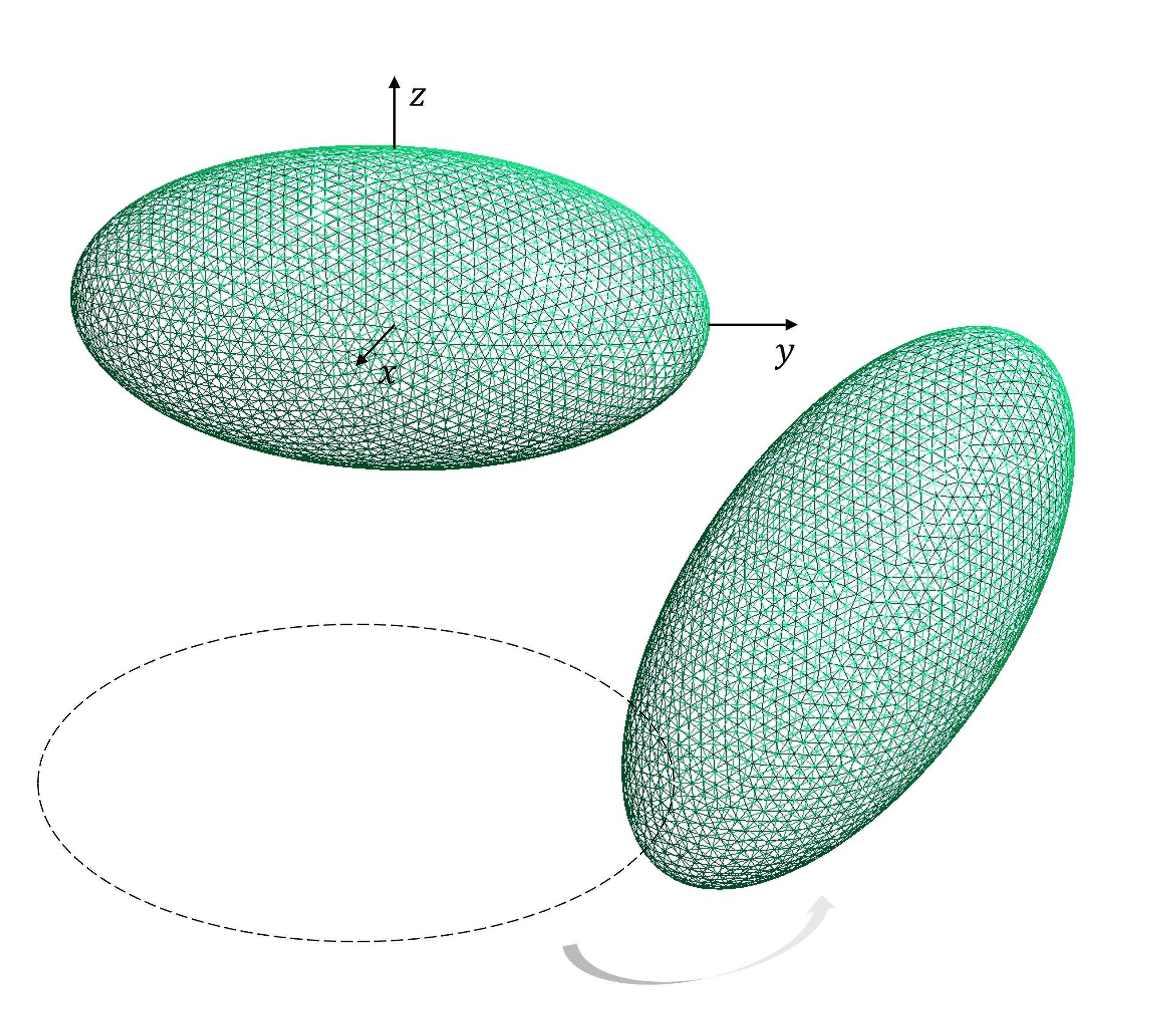}
    \caption{Rotation around x-axis}
    \label{Rotation around x-axis}
    \end{subfigure}
    \caption{Two ellipsoids with or without rotation: when $h_\text{fine}$ = 0.05, $\text{dim}(\mathsf{V}_{\mathrm{i}k}) = 5517$; 
    $h_\text{coarse}$ = 0.1, $\text{dim}(\mathsf{V}_{\mathrm{i}k}) = 1498$. The principal semi-axes of two ellipsoids are $r_{1} = 0.5$ and $r_{2} = 1.0$.}
    \label{Two ellipsoids}
    \end{figure}

\begin{figure}[H]
        \begin{subfigure}{\linewidth}
            \centering
            \includegraphics[scale = 0.5]{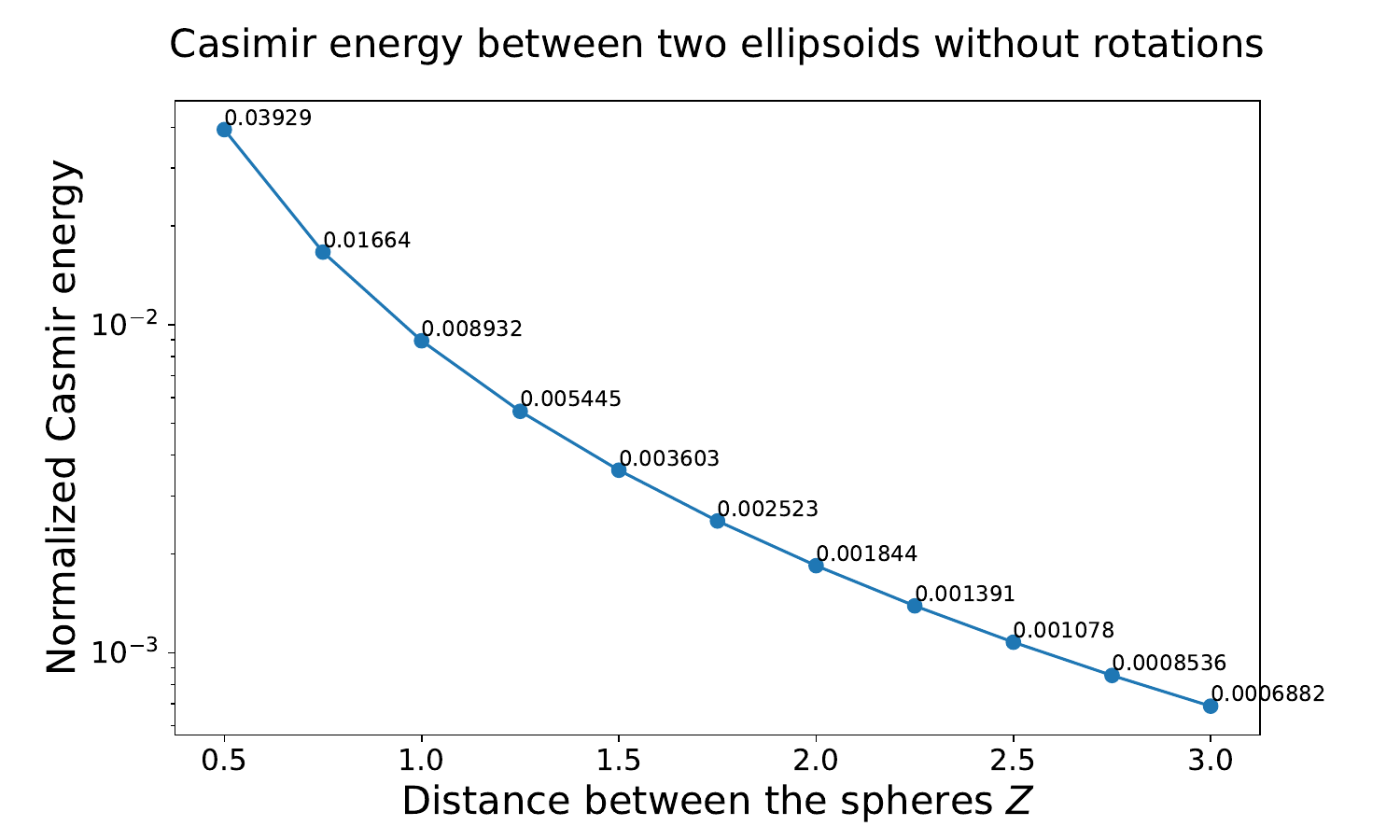}
            \caption{Casimir energy between two ellipsoids with different distances}
            \label{Casimir energy between two ellipsoids with different distances}
            \end{subfigure}\\[1ex]
    
        \begin{subfigure}{\linewidth}
        \centering
        \hspace*{-1cm}\includegraphics[scale = 0.45]{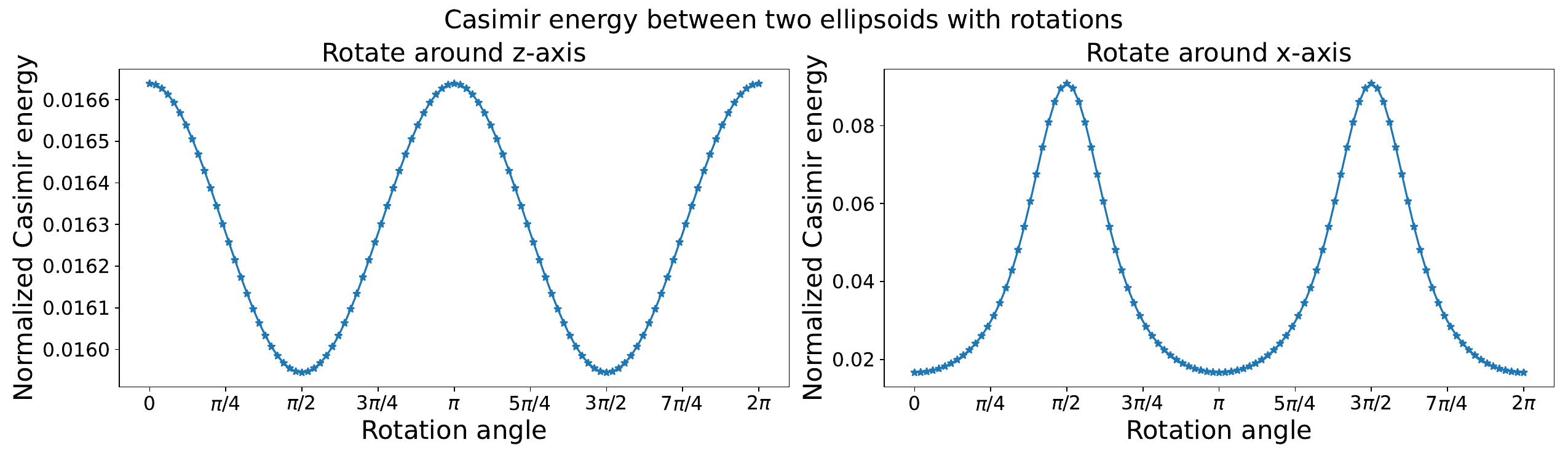}
        \caption{Casimir energy when one of the ellipsoids rotates}
        \label{Casimir energy when one of the ellipsoids rotates}
        \end{subfigure}
        \caption{The dependence of the Casimir energy and rotation angle of one of the ellipsoids.} 
        \end{figure}

In Figure \ref{Without rotation}, the above ellipsoid is centering at $(0,0,0)$ and the below one is centering at $(0, 0, -(0.5+0.5+Z))$, where $Z$ is the 
distance between these two ellipsoids. Without rotation, the Casimir energy between them with different distance $Z$ is plotted in Figure 
\ref{Casimir energy between two ellipsoids with different distances}.

To explore how the rotation affects the change of the Casimir energy, one can always keep one ellipsoid fixed and rotate the other one. The Figure 
\ref{Rotation around z-axis} and \ref{Rotation around x-axis} describe the case when one of the ellipsoids rotates around $z-$ and $x-$axis, respectively.
From the Figure \ref{Casimir energy when one of the ellipsoids rotates}, the Casimir energy changes periodically since we rotate one ellipsoid around 
$z-$ or $x-$axis by 360 degrees.

\begin{figure}[H]
    \begin{subfigure}{\linewidth}
        \centering
        \includegraphics[scale = 0.4]{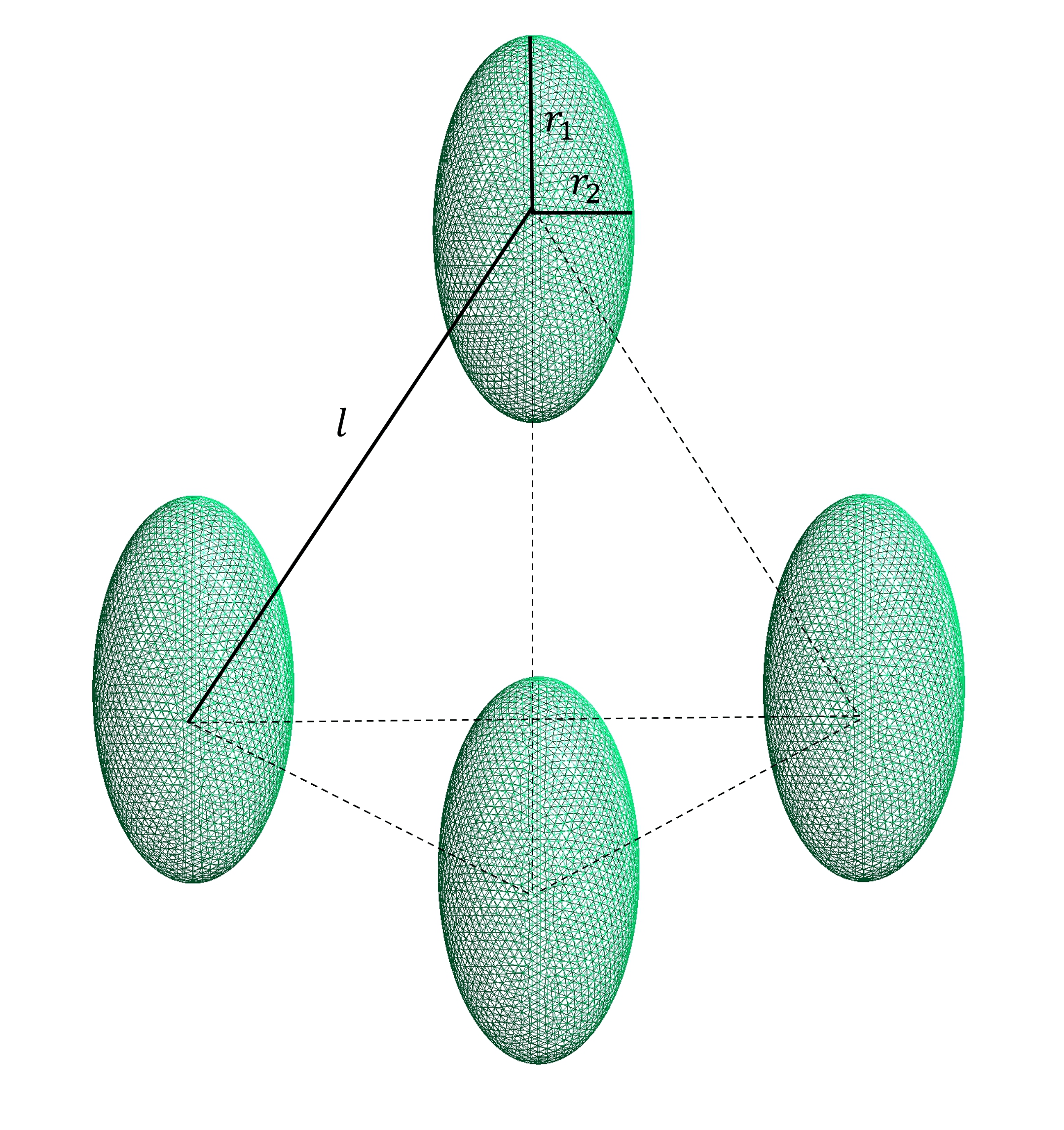}
        \caption{No rotation}
        \label{No rotation 4}
        \end{subfigure}\\[1ex]
    \begin{subfigure}{.5\linewidth}
    \centering
    \includegraphics[scale = 0.09]{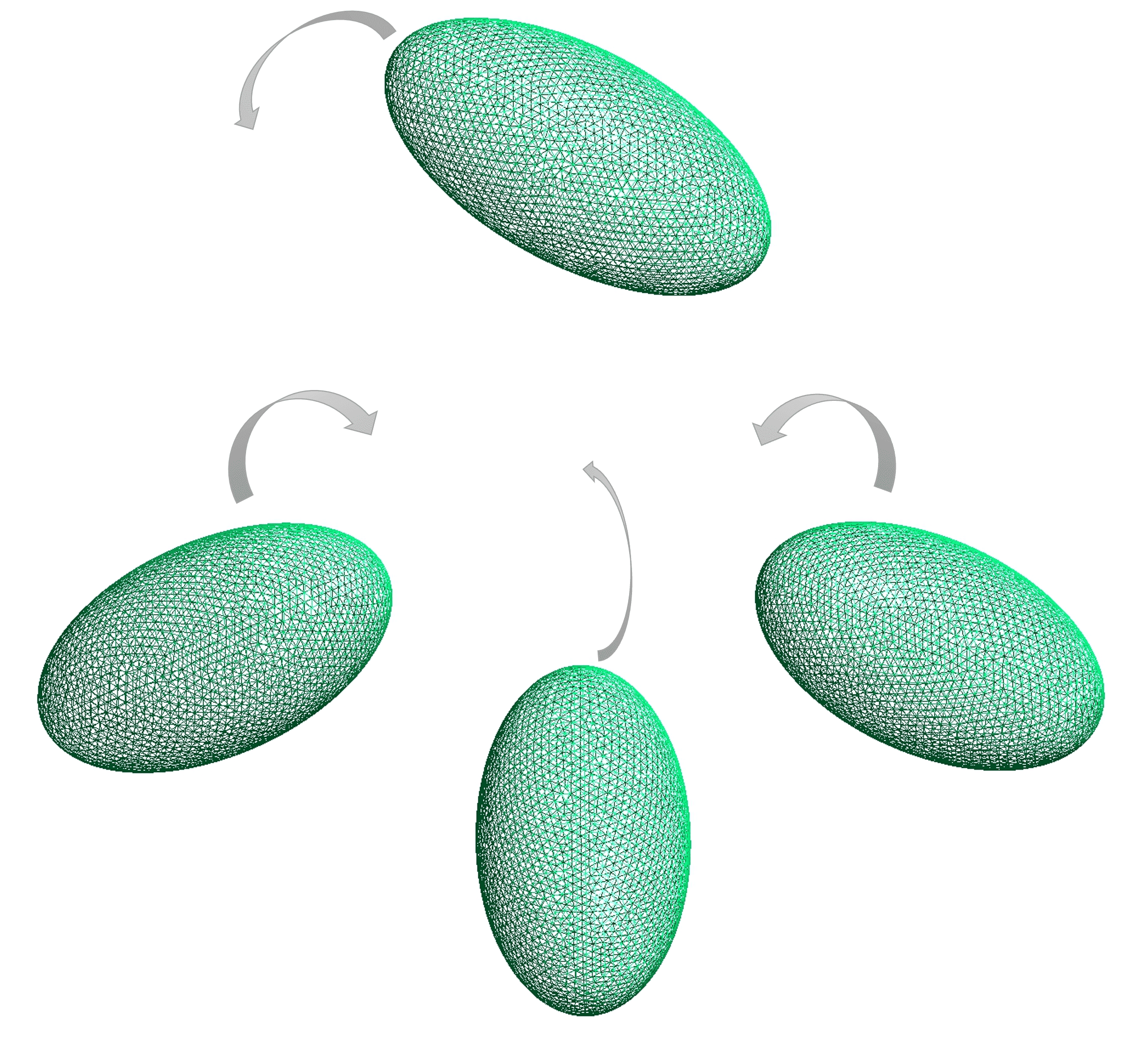}
    \caption{Rotation inwards}
    \label{Rotation inwards 4}
    \end{subfigure}%
    \begin{subfigure}{.5\linewidth}
    \centering
    \includegraphics[scale = 0.4]{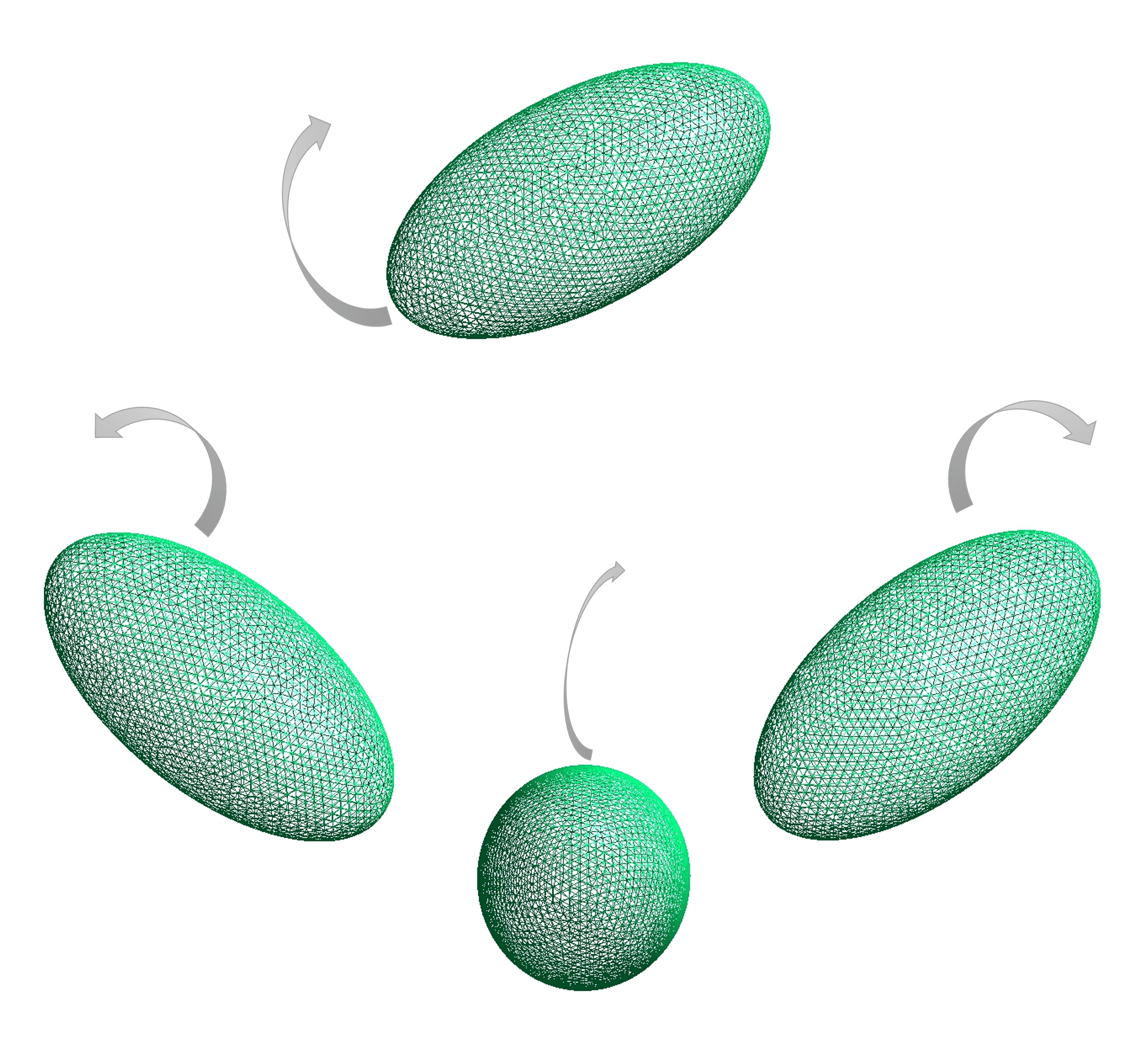}
    \caption{Rotation outwards}
    \label{Rotation outwards 4}
    \end{subfigure}
    \caption{Four ellipsoids with or without rotations:  when $h_\text{fine}$ = 0.03, $\text{dim}(\mathsf{V}_{\mathrm{i}k}) = 11024$; 
    $h_\text{coarse}$ = 0.05, $\text{dim}(\mathsf{V}_{\mathrm{i}k}) = 4160$. The principal semi-axes of these ellipsoids are $r_{1} = 0.6$ and $r_{2} = 0.3$
    and they locate on the vertices of a regular octahedron with edge length $l = 2$.}
    \label{Four ellipsoids with or without rotations}
    \end{figure}

\begin{figure}[H]
        \centering
        \includegraphics[scale = 1]{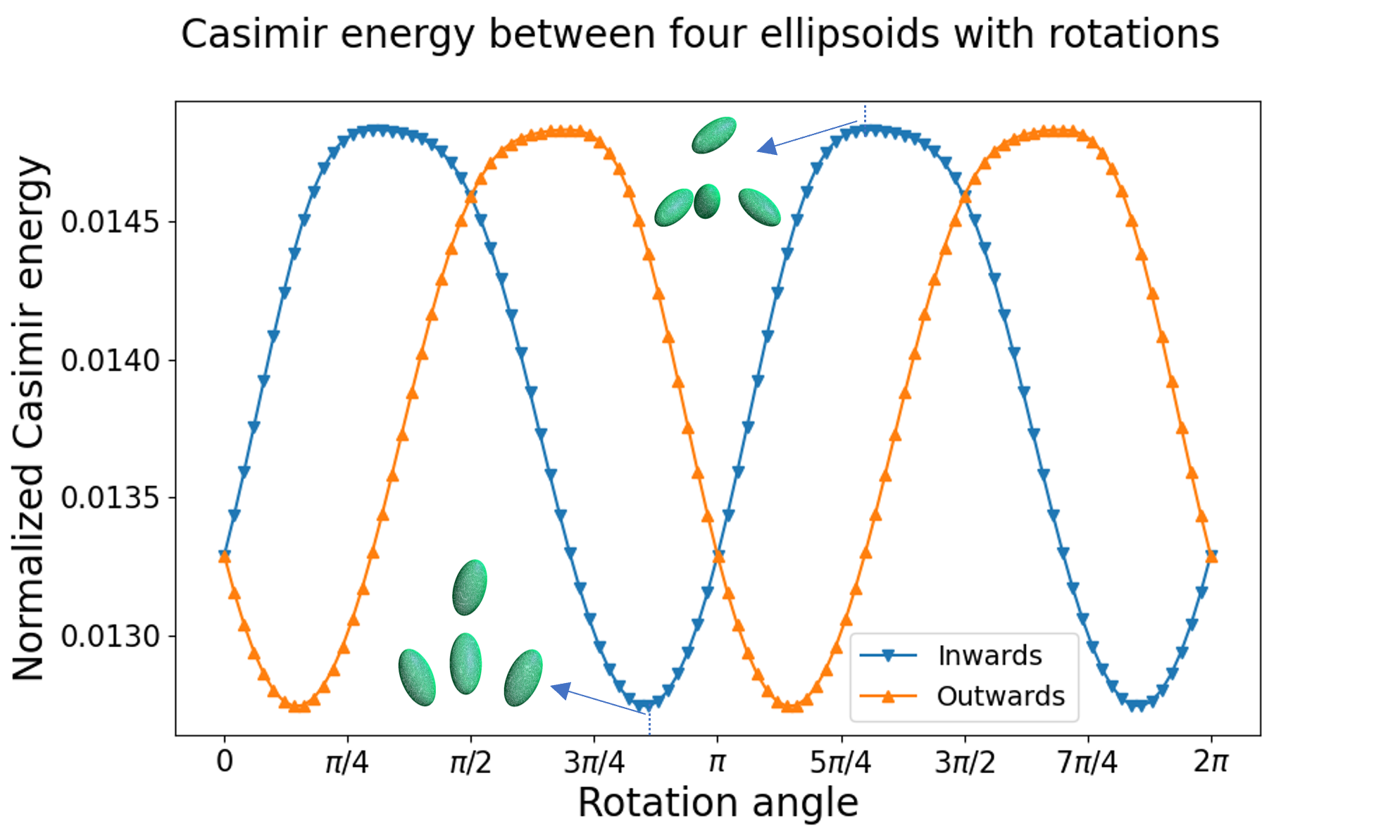}
        \caption{The dependence of the Casimir energy and rotation angle of one of the ellipsoids. Inwards towards the centroid case (solid blue square).
        Outwards towards the centroid case (solid orange triangle).}
        \label{Four ellipsoids}
    \end{figure}

Now, consider 4 ellipsoids located on the vertices of a regular tetrahedron with edge length $l = 2$ (Figure \ref{Four ellipsoids with or without rotations}) and 
the principal semi-axes of all these ellipsoids are $r_{1} = 0.6$ and $r_{2} = 0.3$. Figure \ref{Rotation inwards 4} and Figure \ref{Rotation outwards 4}
show the rotation of the ellipsoids inwards and outwards 360 degrees towards the centroid of this tetrahedron, separately. Afterwards, in order to use the Richardson extrapolation
method to estimate the Casimir energy, we evaluate the integral \eqref{KSSF and CasE} with the grid size set as $h_{\text{fine}} = 0.05$ and 
$h_{\text{coarse}} = 0.03$. Note that the number of the scatterers has increased to four, the matrices $\mathsf{V}_{\mathrm{i}k}$ and 
$\tilde{\mathsf{V}}_{\mathrm{i}k}$ have become to 4 by 4 block and diagonal block matrix, 
respectively. From the Figure \ref{Four ellipsoids}, it shows that the Casimir energy between these four ellipsoids changes periodically with the rotation.

\begin{figure}[H]
    \begin{subfigure}{\linewidth}
        \centering
        \includegraphics[scale = 0.4]{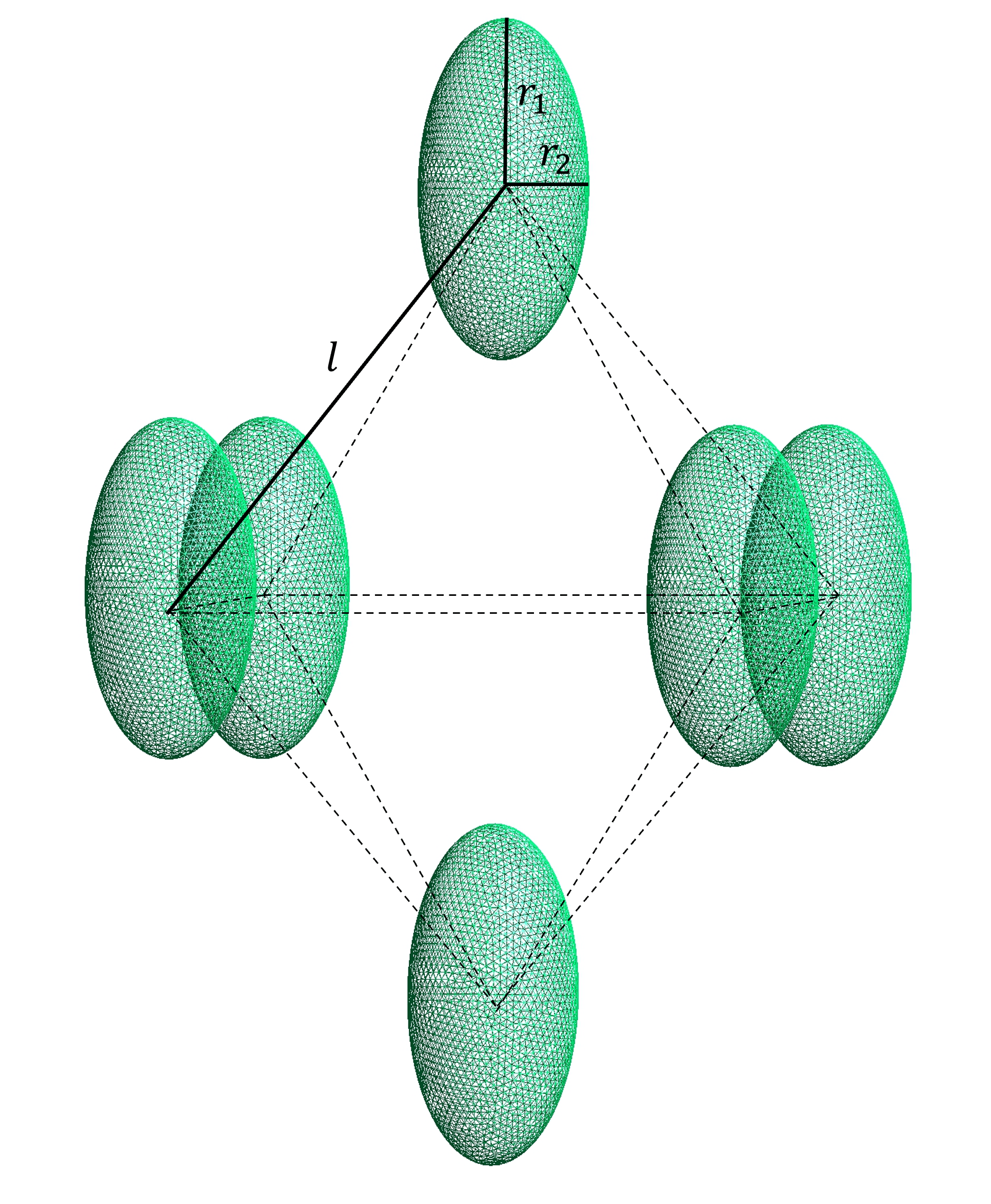}
        \caption{No rotation}
        \label{No rotation 6}
        \end{subfigure}\\[1ex]
    \begin{subfigure}{.5\linewidth}
    \centering
    \includegraphics[scale = 0.4]{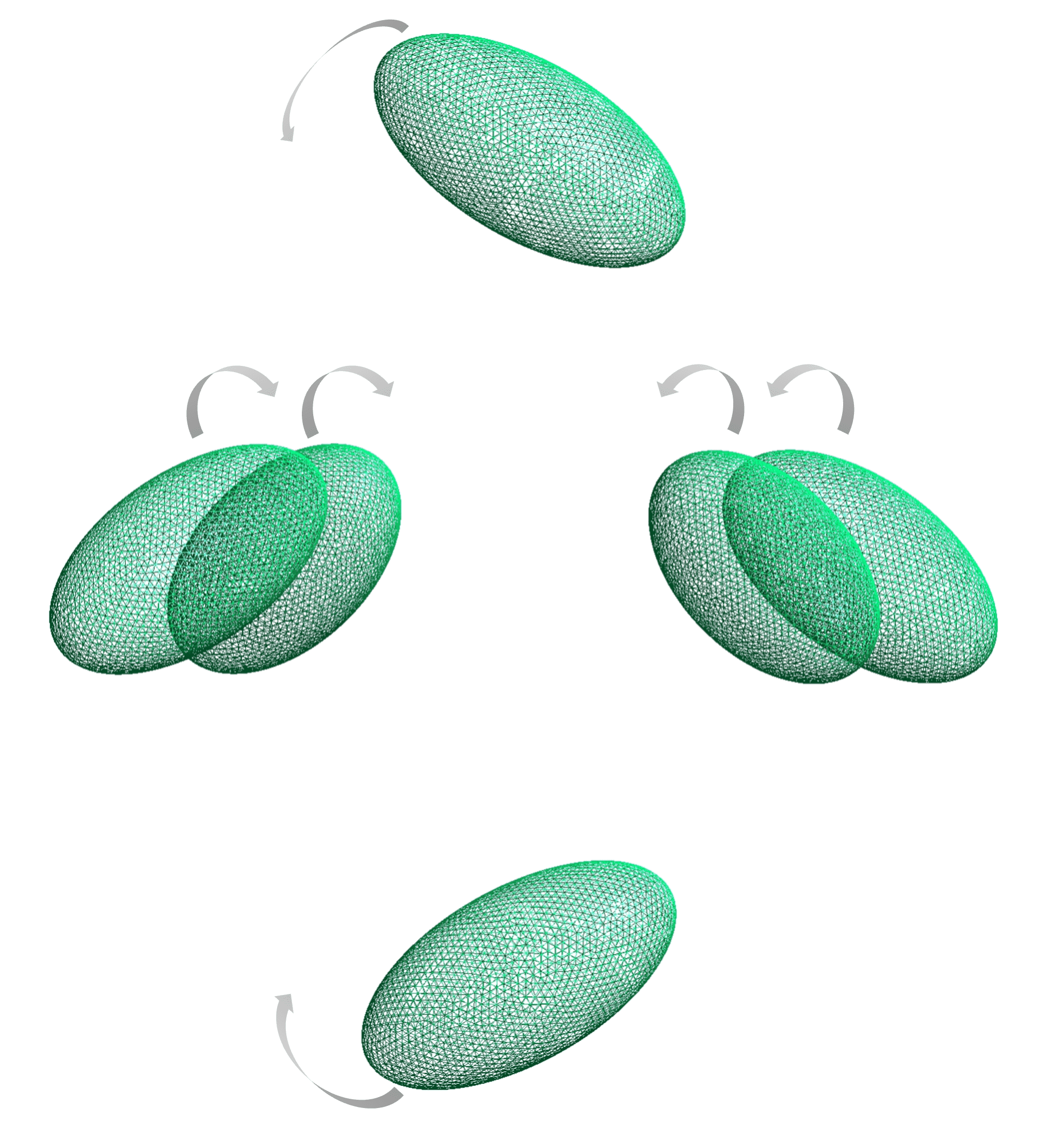}
    \caption{Rotation inwards}
    \label{Rotation inwards 6}
    \end{subfigure}%
    \begin{subfigure}{.5\linewidth}
    \centering
    \includegraphics[scale = 0.4]{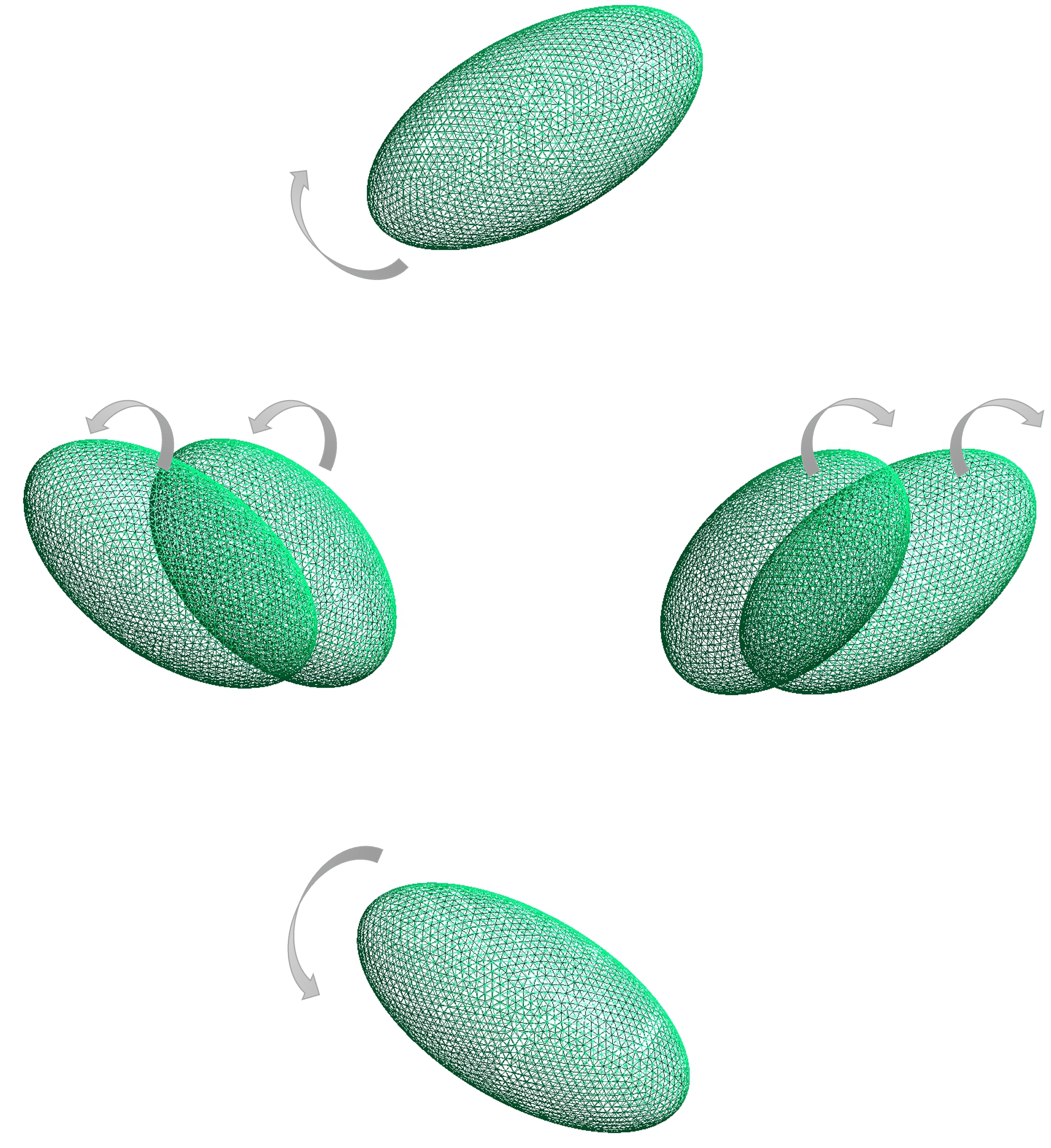}
    \caption{Rotation outwards}
    \label{Rotation outwards 6}
    \end{subfigure}
    \caption{Six ellipsoids with or without rotations: when $h_\text{fine}$ = 0.03, $\text{dim}(\mathsf{V}_{\mathrm{i}k}) = 16536$;  $h_\text{coarse}$ = 0.05, $\text{dim}(\mathsf{V}_{\mathrm{i}k}) = 6240$.
    The principal semi-axes of these ellipsoids are $r_{1} = 0.6$ and $r_{2} = 0.3$
    and they locate on the vertices of a regular octahedron with edge length $l = 2$.}
    \label{Six ellipsoids with or without rotations}
    \end{figure}

\begin{figure}[H]
        \centering
        \includegraphics[scale = 0.5]{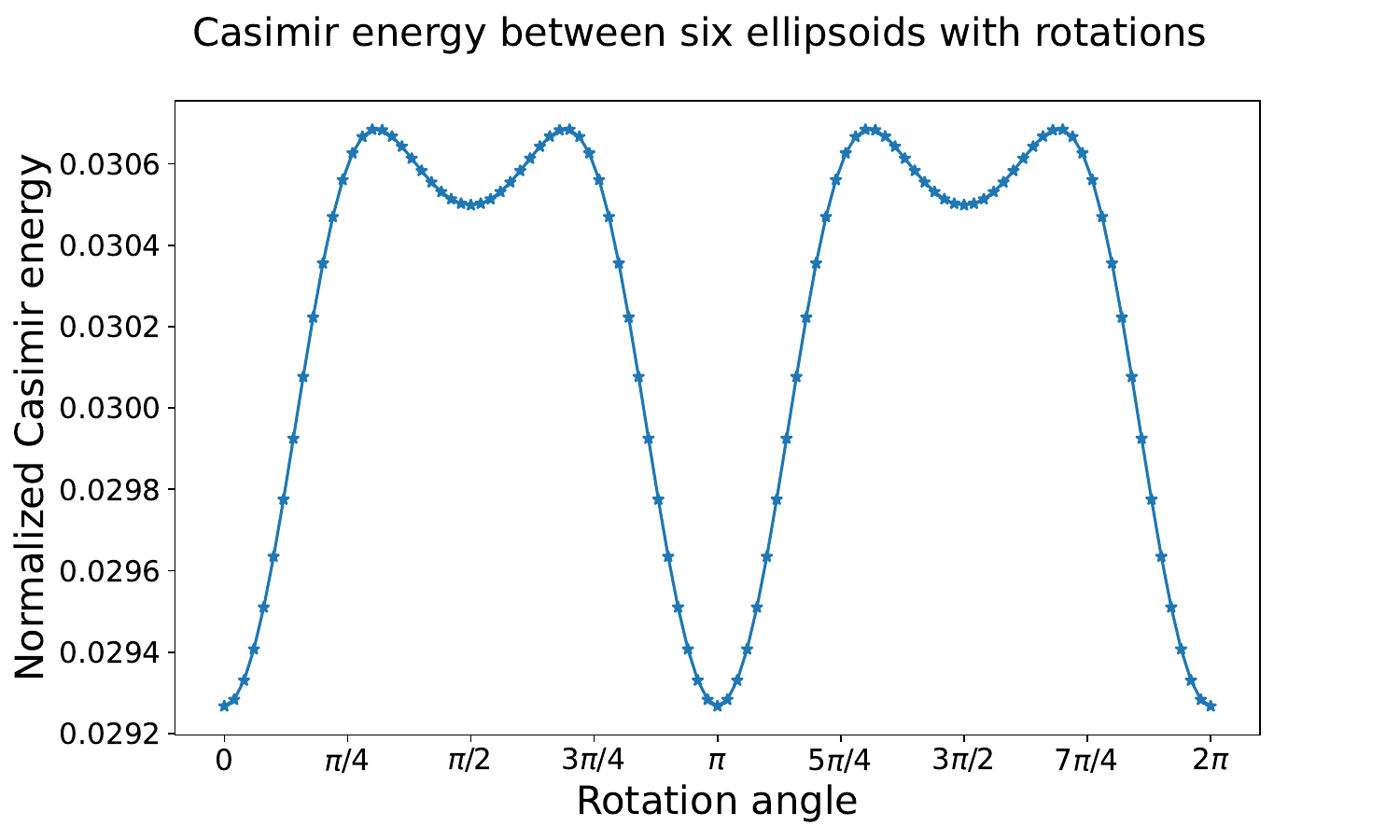}
        \caption{The dependence of the Casimir energy and rotation angle of one of the ellipsoids.}
        \label{Six ellipsoids}
    \end{figure}

  The scatterers of the last example are described inside the Figure \ref{Six ellipsoids with or without rotations}. These six ellipsoids locate on the 
  vertices of a regular octahedron with edge length $l = 2$ and again the principal semi-axes of all these ellipsoids are $r_{1} = 0.6$ and $r_{2} = 0.3$ (shown 
  in the Figure \ref{Six ellipsoids with or without rotations}). This time, the ellipsoids rotate inwards and outwards 360 degrees towards the centroid of 
  this octahedron (Figure \ref{Rotation inwards 6} and Figure \ref{Rotation outwards 6}). By closely looking at these two rotation figures, we can notice that 
  Figure \ref{Rotation inwards 6} can be obtained by rotating Figure \ref{Rotation outwards 6} 180 degrees. Therefore, the Casimir energies for the inwards and 
  outwards cases are the same. Figure \eqref{Six ellipsoids} shows how the Casimir energy changes among these six ellipsoids as they rotate. 

\section{Conclusion}\label{Conclusion}
We have demonstrated in this paper the practical performance and error behaviour for computing the Casimir energy for a number of different configurations, using the log determinant approach. A remaining problem is to speed up this method for large-scale configurations. Here, we demonstrated the performance of different Krylov subspace methods, demonstrating that together with recycling tricks we can significantly reduce the computational effort for large problems.

While we have demonstrated the results in this paper for the acoustic case, the prinicple techniques also transfer to the electromagnetic case. We aim to report corresponding results in future publications.

%\section{Bibliography styles}
% \cite{Feynman1963118}.

% \section*{References}

\bibliography{mybibfile_}

\end{document}